\documentclass[12pt]{article}

\makeatletter\if@twocolumn\PassOptionsToPackage{switch}{lineno}\else\fi\makeatother
\usepackage{graphicx} 
\usepackage[colorlinks=true,linkcolor=blue,citecolor=blue,urlcolor=blue]{hyperref}
\usepackage{listings}
\usepackage{subcaption}
\usepackage{adjustbox}
\usepackage{afterpage}
\usepackage{array}
\usepackage{authblk}
\usepackage{graphicx}
\usepackage[figuresleft]{rotating}
\usepackage{tabularx}
\usepackage{booktabs,multirow}
\usepackage{soul}
\usepackage{amsmath}
\usepackage{amsfonts}
\usepackage{amssymb}
\usepackage{physics}
\usepackage{braket}
\usepackage[utf8]{inputenc}
\usepackage{float}
\usepackage[T1]{fontenc}
\usepackage{url}           
\usepackage[colorinlistoftodos,prependcaption,textsize=tiny]{todonotes}
\usepackage{caption}
\usepackage{threeparttable}
\makeatletter
\edef\fntEncoding{\f@encoding}

\makeatother

\newif\ifmultipleabstract\multipleabstractfalse%
\begin{document}

\pagestyle{myheadings}

\title{Comparative Studies of Quantum Annealing, Digital Annealing, and Classical Solvers for Reaction Network Pathway Analysis and mRNA Codon Selection}

\author[1]{Milind Upadhyay}
\author[1]{Mark Nicholas Jones\thanks{\texttt{research@mqs.dk}}}
\affil[1]{Molecular Quantum Solutions ApS \protect\\ Blegdamsvej 17 \protect\\ 2100 Copenhagen, Denmark}
\maketitle

\begin{abstract}
\noindent
For various optimization problems, the classical time to solution is super-polynomial and intractable to solve with classical bit-based computing hardware to date. Digital and quantum annealers have the potential to identify near-optimal solutions for such optimization problems using a quadratic unconstrained binary optimization (QUBO) problem formulation. This work benchmarks two use cases to evaluate the utility of QUBO solvers for combinatorial optimization problems, in order to determine if a QUBO formulation and annealing-based algorithms have an advantage over classical mixed-integer programming (MIP) and constraint programming (CP) solvers. Various QUBO and solver metrics such as problem mapping, quantitative interconnectivity, penalty structure, solver minimum cost (obtained optimal value) and solver time to solution have been applied to evaluate different QUBO problems. Constrained and unconstrained QUBO solvers are compared including the Fujitsu digital annealer (DA), various D-Wave hybrid quantum annealing solvers (QA, HQA), and the classical MIP/CP solvers HiGHS, Gurobi, SCIP, and CP-SAT. The two industrially relevant use cases are reaction network pathway analysis and mRNA codon selection. For reaction pathway analysis, classical MIP/CP solvers (especially Gurobi and CP-SAT) are observed to solve the problem to optimality in reasonable time frames. For mRNA codon selection, Gurobi outperformed all other solvers in time to solution for all problem sizes, followed by CP-SAT and the D-Wave Nonlinear (NL) HQA solver.
\end{abstract}

\section{Introduction}

Computational solver technology is highly important for industry and society with an impact of up to 4\% energy savings in power distribution \cite{osti_990131}, 25\% reduction in required inventory via inventory optimization for supply chain businesses \cite{idc2009supplychain}, and the prevention of unreliable grids that lead to gross domestic product losses of up to 6\% \cite{iea2023smartgrid}.
\\
Operational research (OR) has a long history of developing solution methods applied to classical combinatorial problems such as the quadratic assignment problem (QAP), traveling salesman problem (TSP), job-shop scheduling and vehicle routing \cite{Kronqvist2025}.
\\
Mixed-integer linear programming (MILP) solvers have profoundly shaped OR, chemical engineering, logistics and other societally relevant industrial sectors and we refer to the scientific literature with respect to such classical computing-based solvers \cite{Clautiaux2025}.
\\
In this work we assess the performance and possible advantage of quantum computing or quantum-inspired solver technology and the following sections will give an introduction to the relevant theory.

\subsection{Binary optimization problems}

A quadratic unconstrained binary optimization (QUBO) problem (isomorphic to an Ising representation) is an optimization problem with terms up to quadratic order.
Binary variables are assigned with values of $q \in \{0, 1\}$ and a cost function for a QUBO can be defined as a sum of linear and quadratic terms:
\begin{equation}
  R(x) = \sum_i Q_{ii} x_i + \sum_{i<j} Q_{ij} x_i x_j
\end{equation}
with $x \in \{0, 1\}^s$ being a vector of binary variables, and $Q \in \mathbb{R}^{s \times s}$ being the QUBO coefficient matrix with scalar quantities for all terms. This cost function is optimized by modifying the variables $x$ through various techniques, to compute the minimum cost and its associated variables $x$ \cite{teplukhin2019qae}.
\\
For instance, a simple QUBO problem with variables $x_1, x_2 \in \{0, 1\}$ can be minimizing the function:
\begin{equation}
  R(x) = -2x_1 + 3x_2 + 4x_1x_2
\end{equation}
The objective of the solver routine is to obtain the minimum cost of $-2$ with $x_1 = 1$ and $x_2 = 0$.
\\
One can extend a QUBO with linear constraints to enforce that the variables encode a feasible solution. For instance, consider a traditionally constrained problem of the form:
\begin{align}
  \text{min } & S(x) = 2x_1 + 3x_2 + -6x_1x_2 \nonumber \\
  \text{s.t. } & x_1 + x_2 \leq 1
\end{align}
where $x_1$ and $x_2$ are binary variables. This constraint allows either or neither variable to be chosen, stopping both from being set to 1. To include such a constraint within a QUBO, one can add a penalty term $\lambda x_1x_2$, where $\lambda$ is a scalar in form of a Lagrange multiplier. Choosing $\lambda$ is a careful balance between ensuring it is large enough to enforce the constraint and small enough to not disturb the objective function; in more complex problems this value often needs to be tuned by an outer optimization loop which can add computational overhead. The QUBO including the constraint is \cite{glover2019tutorialformulatingusingqubo}:
\begin{equation}
  S'(x) = 2x_1 + 3x_2 + -4x_1x_2 + \lambda x_1x_2
\end{equation} 
\noindent
A QUBO can also be extended to a quadratically constrained QUBO (QUBO + QC) with proper separation between the objective function and constraints \cite{fujitsu_da_api}, although the solver routine needs to support additional quadratic constraint equations $P_p(x)$ by enforcing that the variables encode a feasible solution:
\begin{equation}
  P_p(x) = \sum_i C_{ii}^p x_i + \sum_{i<j} C_{ij}^p x_i x_j \leq c^p
\end{equation}
given a constraint definition $C$ having $|C|$ different equations, with each equation $C^p$ having a comparison operator and an associated constant $c^p$. Each constraint essentially enforces that a quadratic equation containing certain variables is less than or equal to a constant; such equations can also be written as equality constraints. The solver will ensure solution $x$ is feasible, or satisfying all constraints, before minimizing the cost function.
\\
Higher-order binary optimization (HOBO) problems are a generalization of QUBOs with terms higher than quadratic order. Such a cost function can be defined as:
\begin{equation}
  R(x) = \sum_i Q_{i} x_i + \sum_{i<j} Q_{ij} x_i x_j + \sum_{i<j<k} Q_{ijk} x_i x_j x_k + \cdots
\end{equation}
where the sum can go up to an arbitrary order (cubic, quartic, etc.), with $x \in \{0, 1\}^s$ being a vector of binary variables, and $Q \in \mathbb{R}^{s \times s \times s \times \cdots}$ being the HOBO coefficient matrix.
\\
One can also extend a HOBO to a constraint-based HOBO with constraints such as:
\begin{equation}
  P_p(x) = \sum_i C_{i}^p x_i + \sum_{i<j} C_{ij}^p x_i x_j + \sum_{i<j<k} C_{ijk}^p x_i x_j x_k + \cdots \leq c^p
\end{equation}
given a constraint definition $C$ having $|C|$ different equations, with each equation $C^p$ having a comparison operator and an associated constant $c^p$.

\subsection{Overview of computational solvers}

IPOPT is an open-source solver specialized in nonlinear programming (NLP), effectively finding local optima for nonlinear, continuous problems, including linear programming (LP), quadratic programming (QP), and quadratic programming with quadratic constraints (QP + QC) \cite{wachter2006implementation}. HiGHS is an open-source solver that can solve LP/QP and mixed-integer linear programming (MILP) problems \cite{huangfu2018parallelizing}. SCIP is an open-source solver which combines constraint programming (CP) with MILP, mixed-integer quadratic programming (MIQP), and mixed-integer nonlinear programming (MINLP) problems \cite{BolusaniEtal2024OO}. Constraint Programming Satisfiability (CP-SAT) is an open-source CP solver \cite{cpsatlp}. Gurobi is a commercial solver that can solve MILP, MIQP, MINLP, quadratic unconstrained binary optimization (QUBO) and those with quadratic constraints (QUBO + QC), as well as continuous optimization problems \cite{gurobi}.
\\
For quantum and quantum-inspired optimization, the Fujitsu Digital Annealer (DA) for example is tailored for binary optimization problems, supporting QUBO and QUBO + QC problems \cite{katayama2024}. In comparison, D-Wave’s Quantum Annealer (QA) similarly addresses QUBO problems but lacks native constraint support, though the D-Wave Leap CQM Hybrid Quantum Annealing solver (HQA) extends capabilities with a combination of classical optimization techniques and quantum annealing, directly handling constraints and also supporting MILP, MIQP, LP/QP (and with QC), and QUBO + QC problems. The Leap Nonlinear (NL) HQA also supports HOBO and MINLP problems as well as CP formulations with decision variables \cite{dwave_systems}.
The capabilities of these solvers are summarized in Table \ref{tab:solver_capabilities}.

\begin{table}[H]
  \centering
  \begin{threeparttable}
    \caption{Solver capabilities across optimization problem types; DA: Digital Annealer; QA: Quantum Annealer; HQA: Hybrid Quantum Annealer}
    \label{tab:solver_capabilities}
    \begin{tabularx}{\linewidth}{|>{\centering\arraybackslash}X|>{\centering\arraybackslash}X|>{\centering\arraybackslash}X|>{\centering\arraybackslash}X|>{\centering\arraybackslash}X|>{\centering\arraybackslash}X|>{\centering\arraybackslash}X|>{\centering\arraybackslash}X|>{\centering\arraybackslash}X|}
      \hline
      Type    & IPOPT       & HiGHS       & SCIP        & CP-SAT      & Gurobi                   & DA          & QA          & HQA         \\
      \hline
      LP              & \checkmark  & \checkmark  &             &             & \checkmark               &             &             &   \checkmark          \\
      \hline
      QP              & \checkmark  & \checkmark  &             &             & \checkmark               &             &             &   \checkmark          \\
      \hline
      QP + QC         & \checkmark  &             &             &             & \checkmark               &             &             &     \checkmark        \\
      \hline
      NLP             & \checkmark  &             &             &             &   
              &             &             &             \\
      \hline
      MILP            &             & \checkmark  & \checkmark  &             & \checkmark               &             &             & \checkmark  \\
      \hline
      MIQP            &             &             & \checkmark  &             & \checkmark               &             &             & \checkmark  \\
      \hline
      MINLP           &             &             & \checkmark  &             & \checkmark               &             &             &   \checkmark       \\
      \hline
      QUBO            &             &             & \checkmark  &             & \checkmark               & \checkmark  & \checkmark  & \checkmark  \\
      \hline
      QUBO + QC       &             &             & \checkmark  &             & \checkmark               &     \checkmark        &             & \checkmark  \\
      \hline
      HOBO            &             &             & \checkmark  &             & \checkmark               &   &  &   \checkmark       \\
      \hline
      CP              &             &             & \checkmark  & \checkmark  &                          &             &             &      \checkmark       \\
      \hline
    \end{tabularx}
  \end{threeparttable}
\end{table}
\noindent
Note that while the DA does not natively support solving HOBO problems, it does provide tools for converting HOBO problems to QUBO problems (that can then be solved) via Fujitsu's Digital Annealer API \cite{fujitsu_da_api}.

\subsection{Simulated, quantum-inspired, and quantum annealing}

\subsubsection{Simulated annealing}
Simulated annealing (SA) is a classical optimization technique that explores a solution landscape with techniques inspired by physical annealing processes in metallurgy \cite{mukherjee2015}.
\\
SA initializes a system for an optimization problem in a high-temperature state (high noise in adjusting the problem variables) to access the whole range of the cost function,
and then gradually lowers the temperature of the optimization to ideally reach the global minimum of the cost function.
The simulated thermal noise allows the system to escape local minima in the cost function.
However, when energy barriers between local minima and the global minimum are substantially high, SA fails to converge efficiently and the search time grows with $O(e^N)$ for a problem size of $N$ \cite{mukherjee2015}.
\\
In SA, the cost function to optimize with variables $x$ is referred to as an energy function $E(x)$ and represented as a Hamiltonian $H(x)$ \cite{ingber1993simulated}
\begin{equation}
  H(x) = E(x)
  \label{eq:sa_hamiltonian}
\end{equation}
\\
For instance, in a QUBO problem, the Hamiltonian for SA is the equation
\begin{equation}
  H(x) = \sum_i Q_{ii} x_i + \sum_{i<j} Q_{ij} x_i x_j
  \label{eq:sa_qubo_hamiltonian}
\end{equation}
where $Q$ is the QUBO matrix \cite{padmasola2025solvingtravelingsalesmanproblem}.
\\
At a given temperature \(T\) during SA, a modification to the variables that changes the cost by \(\Delta E\) from the previous cost is accepted with a Metropolis acceptance rule with probability
\begin{equation}
P_{\mathrm{SA}}(\Delta E)\;=\;\frac{1}{1+\exp(\Delta E/T)}
          \simeq \exp(-\Delta E/T)
\label{eq:sa_metropolis}
\end{equation}
which is a Boltzmann distribution and guarantees detailed balance that all possible states of the system can theoretically be sampled \cite{ingber1993simulated}. It can be seen from this equation how the probability of accepting a change to a state with a higher cost decreases with decreasing temperature.

\subsubsection{Quantum-inspired annealing}
Various quantum-inspired annealers have been developed as alternatives to simulated annealing with potentially improved performance without the need for reliable quantum hardware, including simulated coherent Ising machines, simulated bifurcation machines, and digital annealers that simulate mechanisms inspired by quantum processes including entanglement and superposition, with technology such as pulsed lasers, FPGAs, GPUs, etc. \cite{zeng2024qaia, katayama2024, kashimata2024efficient}. These methods are heuristic in nature similar to simulated annealing and do not guarantee to find the global minimum of the cost function.
\\
For instance, a Digital Annealer (DA) minimizes a QUBO cost function via a parallelized Metropolis acceptance rule.
\\
The Hamiltonian for a DA, $H(x)$, represents the cost function which is also referred to as the energy function. It is the same as the SA Hamiltonian for a QUBO problem as given in equation \eqref{eq:sa_qubo_hamiltonian}.
\\
For a single bit flip in the vector of binary variables $x$ with respect to QUBO $Q$, the local field
\begin{equation}
h_i=\sum_j Q_{ij}x_j + Q_{ii}
\end{equation}
gives the energy change as a sum of all the terms in the QUBO that are affected by the flip of the $i$th bit. The energy change based on $\Delta x_i \in \{-1, 1\}$ is therefore
\begin{equation}
\Delta E_i \;=\; \Delta x_i\,h_i .
\end{equation}  
The probability of accepting a certain bit flip given temperature $T$ follows a Metropolis distribution similar to SA,
\begin{equation}
P_{\mathrm{DA}}(\Delta E_i)\;=\;\min\!\bigl[1,\exp(-\Delta E_i/T)\bigr]
\label{eq:da_metropolis}
\end{equation}
\\
Given the nature of additions/subtractions applied during this search, the DA can be parallelized with GPUs across all bits and for multiple different temperature configurations, facilitating escaping local minima via the parallel tempering algorithm \cite{katayama2024}.
\\
Equations \eqref{eq:sa_metropolis} and \eqref{eq:da_metropolis} show the common Metropolis acceptance probability between SA and DA, with higher temperatures leading to easier acceptance of higher cost states. DA differs from SA in its native support for parallelization in this optimization, with inspiration from quantum annealing in terms of its functionality for parallelization and escaping local minima.

\subsubsection{Quantum annealing}
Quantum annealing (QA) is an analog quantum computing based optimization routine. QA aims to solve the SA convergence problem with quantum fluctuations, effectively utilizing tunneling to escape local minima that have high but thin barriers. In QA, the search time for these certain problems grows with $O(e^{\sqrt N})$, thus scaling more efficiently than simulated annealing for which the search time on corresponding problems grows with $O(e^N)$ \cite{mukherjee2015}.
\\
While SA and DA explicitly evaluate discrete bit flips through a classical Metropolis test as seen in equations \eqref{eq:sa_metropolis} and \eqref{eq:da_metropolis}, QA lets qubits evolve simultaneously in an analog manner under a time-dependent Hamiltonian:
\begin{equation}
  H(t) = (1 - s(t)) H_0 + s(t) H'
  \label{eq:adiabatic_evolution}
\end{equation}
where $H_0$ is the Hamiltonian describing the cost function to minimize (similar to the Hamiltonian $H(x)$ employed in SA and DA), $H'$ is a Hamiltonian with a known ground state, and $s(t)$ is a function used to interpolate between the two Hamiltonians.
\\
At the start ($s(t_0) = 1$), the quantum system is initialized at the known ground state of $H'$.
As time evolves, $s(t)$ decreases such that the cost function Hamiltonian $H_0$ becomes the dominant term with $s(t_{tot}) = 0$ at the end of the annealing process.
\\
If the total annealing time $t_{tot}$ is long enough (for a more gradual decrease of $s(t)$), the system evolves adiabatically with the instantaneous Hamiltonian $H(t)$.
At $t$ the system remains at the ground state of $H(t)$ according to the quantum adiabatic theorem \cite{mukherjee2015,king2023qaspinglass}.
Under these ideal conditions, at the end of the annealing time ($H(t_{tot}) = H_0$) the system has converged to the ground state of $H_0$ which corresponds to the optimal solution of the cost function.
\\
Similar to how SA and DA lower the system temperature $T$ over time to precisely optimize around the final solution, QA decreases $s(t)$ over time to approach the ground state of the cost function Hamiltonian $H_0$ and obtain its final solution.
\\
In practice, optimal annealing times can be significantly longer than qubit coherence times, leading to diabatic transitions in quantum annealing and suboptimal solutions \cite{weinberg2020scaling, king2023qaspinglass}.
\\

\subsubsection{Quantum adiabatic theorem}

The quantum adiabatic theorem is a fundamental concept in quantum mechanics that dictates the time evolution of quantum systems under slowly varying conditions. Originally formulated by Born and Fock in 1928 \cite{born1928adiabatic}, the theorem states that a quantum system initialized in an eigenstate of a time-dependent Hamiltonian will remain in the corresponding instantaneous eigenstate throughout the evolution, provided the Hamiltonian changes sufficiently slowly. This contrasts with diabatic processes, where faster changes in the system can cause transitions between different energy eigenstates, leading to excitations and the system becoming a linear combination of eigenstates. The adiabatic theorem was later given a more rigorous mathematical proof by Kato \cite{kato1950adiabatic}. 
\\
The key condition for adiabatic evolution is that the rate of change of the Hamiltonian must be much smaller than the square of the minimum energy gap between the instantaneous ground state and the first excited state. When this gap becomes small, the adiabatic condition requires considerably higher annealing times \cite{suzuki2022quantum}. Formally, the minimum gap requirement is defined as:

\begin{equation}
\frac{\max \|\langle E_1(t)|(dH(t)/dt)|E_0(t)\rangle\|}{\min |\Delta(t)|^2} \ll 1,
\end{equation}
\noindent
where $H(t)$ is the system's Hamiltonian at time $t$, $|E_0(t)\rangle$ and $|E_1(t)\rangle$ are at time $t$ the ground energy state and first excited state, and $\Delta(t)$ is the energy gap at time $t$ between $|E_0(t)\rangle$ and $|E_1(t)\rangle$. The minimum and maximum are obtained with respect to time $t$ \cite{suzuki2022quantum}.
The different annealing methods are summarized in Table \ref{tab:annealing_paradigm_comparison}.

\begin{table}[H]
  \centering
  \footnotesize
  \begin{threeparttable}
    \caption{Comparison of annealing paradigms}
    \label{tab:annealing_paradigm_comparison}
    \begin{tabularx}{\linewidth}{|>{\centering\arraybackslash}X|>{\centering\arraybackslash}X|>{\centering\arraybackslash}X|}
      \hline
      \textbf{Paradigm} & \textbf{Optimization method}                     & \textbf{Optimality} \\
      \hline
      Simulated         & Simulates thermal fluctuations                   & Heuristic           \\
      \hline
      Quantum-Inspired  & Simulates quantum effects & Heuristic           \\
      \hline
      Quantum           & Quantum tunneling, adiabatic evolution           & Heuristic           \\
      \hline
    \end{tabularx}
  \end{threeparttable}
\end{table}

\subsection{Commercial and open-source quantum and quantum-inspired annealing solvers}

In terms of quantum-inspired computing, Fujitsu's quantum-inspired classical computing digital annealer (DA) is one of the available solver technologies inspired by QA. The Fujitsu DA v4 applies high-performance computing (HPC) with GPUs, utilizing parallelism and connectivity between all bit variables in optimization problems with the intention of achieving effects similar to quantum superposition, tunneling, and entanglement \cite{katayama2024}.
\\
Toshiba's simulated bifurcation machine (SBM) is a quantum-inspired device using GPU/FPGA-based computing to achieve similar quantum effects \cite{zeng2024qaia, kashimata2024efficient}.
\\
At the time of writing (August 2025), D-Wave offers commercial quantum annealing solvers, such as Advantage and Advantage2, which are superconducting qubit-based quantum annealers with the newest models containing 5000 qubits \cite{dwave_systems}. Strong evidence for coherent dynamics in these quantum annealers at short timescales such as 30 nanoseconds has been presented. In practice, many problems require significantly longer anneal times, for which the behavior is expected to depart from coherence and involve thermal effects \cite{king2023qaspinglass}; tunneling is expected to be mostly incoherent and long-lived entanglement is unlikely.
\\
Open-source quantum annealing development frameworks have also been developed, such as QuantRS2-Anneal \cite{quantrs2_anneal} and GPU-pSAv \cite{onizawa2025gpu}. QuantRS2-Anneal is a toolkit with various quantum and quantum-inspired annealing processes that can be simulated locally or run on the cloud (D-Wave, AWS Braket, Fujitsu DA). GPU-pSAv is a quantum-inspired annealing solver built on probabilistic bit (p-bit) based simulated annealing that can be run on GPUs.
\\
Table \ref{tab:annealing_solvers} shows a comparison of commercial and open-source quantum annealing (QA) and quantum-inspired (QI) solvers from various companies for which there have been several studies published comparing their performance.
\begin{table}[H]
  \footnotesize
  \centering
  \begin{threeparttable}
    \caption{Commercial and open-source annealing solvers;\\ QA: quantum annealing, QI: quantum inspired, SDK: software development kit}
    \label{tab:annealing_solvers}
    \begin{tabularx}{\linewidth}{|>{\centering\arraybackslash}X|>{\centering\arraybackslash}X|>{\centering\arraybackslash}X|>{\centering\arraybackslash}X|>{\centering\arraybackslash}X|}
      \hline
      \textbf{Company / } & \textbf{Solver Name}          & \textbf{Paradigm} & \textbf{Hardware} & \textbf{Ref.} \\
      \textbf{Institute} &  & &  & \\
      \hline
      D-Wave           & Advantage, Advantage2          & QA                & Superconducting Qubits   & \cite{dwave_systems}   \\
      \hline
      Fujitsu          & Digital Annealer              & QI                & CPU, GPU    & \cite{katayama2024} \\
      \hline
      Toshiba          & Simulated Bifurcation & QI                & CPU, GPU, FPGA    & \cite{kashimata2024efficient} \\
      &              Machine         &                   &                   & \\
      \hline
      COOLJAPAN OÜ     & QuantRS2-Anneal               & QA \& QI SDK          & Any          & \cite{quantrs2_anneal} \\
      \hline
      Tohoku University Electrical Communication     & GPU-pSAv                      & QI                & CPU, GPU          & \cite{onizawa2025gpu} \\
      \hline
    \end{tabularx}
  \end{threeparttable}
\end{table}
\noindent
Various work has compared the performance of QA to SA and the quantum-inspired solvers and the following observations, conclusions and claims have been made:
\begin{itemize}
  \item D-Wave's superconducting qubit quantum annealers (Advantage and Advantage2) achieved better performance than leading classical methods (matrix product states, projected entangled pair states, neural quantum states) for simulating quenched dynamics of spin glasses in 2D, 3D, and infinite-dimensional systems. \cite{king2025beyond}. The classical computational methods evaluated are distinct from annealing approaches.
  \item QA on D-Wave's 2000 and 5000-qubit quantum annealers has outperformed SA for the maximum cardinality problem, in which the embedding of optimization problems to the quantum computer's architecture was shown to be important for performance \cite{vert2024qamaxcardinality}.
  \item In problems such as the max-cut problem, quantum-inspired algorithms have been shown to outperform QA \cite{zeng2024qaia}.
  \item The performance of D-Wave's HQA utilizing both quantum annealing and classical processing has also been compared to Fujitsu's DA and Toshiba's SBM, with HQA outperforming the other methods on MQLib problem instances (some from real-world problems), DA outperforming the others on the random not-all-equal 3-SAT, and the SBM outperforming the others on the Ising spin glass Sherrington-Kirkpatrick (SK) model \cite{oshiyama2022qis}; it is interesting to note that in this case, D-Wave's hybrid QA Solver showed suboptimal performance in the spin-glass problem compared to the SBM.
  \item Pfizer partnered with D-Wave and QuantumBasel to enhance production scheduling for their Freiburg plant for reducing energy consumption and improving capacity planning, benchmarking a classical GPU-based solver and the D-Wave Leap NL HQA solver and observing the HQA solver to slightly outperform the classical solver \cite{quantumbasel2025pfizer}.
  \item An older version of the Fujitsu DA, based on application-specific CMOS hardware, was benchmarked against single-core methods: SA and parallel tempering with isoenergetic cluster moves (PT + ICM). For sparse two-dimensional spin-glass problems the DA did not demonstrate improved efficiency, but for spin-glass problems with full connectivity the DA demonstrated a speedup of around 2 orders of magnitude in time to solution \cite{aramon2019physics}.
  \item For chemical reaction network pathway analysis, simulated annealing and the D-Wave Advantage QA were benchmarked against the classical Gurobi solver, finding that the computational time to an optimal reaction pathway scaled more inefficiently with SA and QA \cite{mizuno2024optimal}.
  \item For mRNA codon optimization, the D-Wave HQA and quantum approximation optimization algorithm (QAOA) on a Qiskit simulator were benchmarked against a classical genetic algorithm, with neither of the HQA or QAOA methods being able to outperform the genetic algorithm \cite{fox2021}.
\end{itemize}
Table \ref{tab:annealing_use_cases} summarizes benchmarked use cases from the above literature with the different annealing methods.

\begin{table}[H]
  \centering
  \begin{threeparttable}
    \caption{Overview of use cases and which annealing methods (and hybrid schemes) have been applied.}
    \label{tab:annealing_use_cases}
    \begin{tabularx}{\linewidth}{|>{\centering\arraybackslash}X|>{\centering\arraybackslash}X|>{\centering\arraybackslash}X|>{\centering\arraybackslash}X|}
      \hline
      \textbf{Problem type}                         & \textbf{SA} & \textbf{DA/SBM} & \textbf{QA/HQA} \\
      \hline
      Max cardinality                               & \checkmark  &                 & \checkmark  \\
      \hline
      Spin-glass systems                            & \checkmark  & \checkmark      & \checkmark  \\
      \hline
      MQLib                                         &             & \checkmark      & \checkmark  \\
      \hline
      Random not-all-equal 3-SAT                    &             & \checkmark      & \checkmark  \\
      \hline
      Max-cut                                       &             & \checkmark      & \checkmark  \\
      \hline Production scheduling                  &             &                 & \checkmark  \\
      \hline
      Reaction Network Pathway Analysis             & \checkmark  &                 & \checkmark  \\
      \hline
      mRNA Codon Optimization                       &             &                 & \checkmark  \\
      \hline
    \end{tabularx}
  \end{threeparttable}
\end{table}
\noindent
A takeaway from these different benchmark studies is that one must carefully evaluate the energy landscape of their optimization problem and test/benchmark different annealing techniques since the performance between the different methods is problem-dependent.
\\
Marthaler et al. \cite{marthaler2025goodusecasequantum} have developed a framework for determining useful applications of quantum computing, which includes identifying real-world problems, assessing their mappings to quantum computers, solving the problems classically, and demonstrating quantum utility. It focuses on evaluating how likely a problem is to have practical utility with quantum computing, in contrast to this work that focuses on evaluating the most effective solver approach for a given problem (quantum/quantum-inspired versus classical) based on QUBO problem structure metrics.

\section{Metrics for QUBO benchmark framework}
\label{sec:qubo_metrics}
This work presents a benchmarking framework for QUBO-based optimization problems that can assist in determining the possible utility of using a quantum-based QUBO solver for a specific optimization problem.
\\
The benchmarking framework considers the following metrics:
\begin{enumerate}
  \item  Problem Mapping Metrics
        \begin{itemize}
          \item Pre-processing problem to QUBO matrix
                \begin{itemize}
                  \item Case A) superlinear bloating to QUBO from binarization
                  \item Case B) linear scaling to QUBO from binarization
                  \item Case C) logarithmic scaling to QUBO from binarization
                  \item Case D) one-to-one mapping to QUBO from binarization
                  \item Case E) HOBO to QUBO mapping inefficiency
                \end{itemize}
          \item Post-processing solution improvement complexity
        \end{itemize}
  \item QUBO Analysis Metrics
        \begin{itemize}
          \item Metric I: Quantitative connectivity
          \begin{itemize}
            \item Size
            \item Density
            \item Interconnectivity
            \item Rank-1 dominance
          \end{itemize}
          \item Metric II: Penalty structure
                \begin{itemize}
                  \item Constraint type
                  \begin{itemize}
                    \item Case A) Linear
                    \item Case B) Quadratic
                    \item Case C) One-hot
                  \end{itemize}
                  \item Penalty separation
                \begin{itemize}
                  \item Case A) embedded in cost function, need parameter tuning
                  \item Case B) separated into penalty QUBO
                  \end{itemize}
                \end{itemize}
        \end{itemize}
  \item Solver performance metrics
        \begin{itemize}
          \item Minimum cost obtained
          \item Time to solution
        \end{itemize}
\end{enumerate}

\subsection{Problem mapping metrics}
One key factor in determining whether to solve a problem with a QUBO solver is the efficiency of pre-processing the problem matrix to the QUBO matrix mappings, or the creation of a QUBO matrix given a natural representation of the problem. Some problems are naturally represented with binary variables, for which there is a one-to-one mapping to variables in a QUBO. However, often one has to encode integers or real numbers into binary (binarization) for the QUBO (unary, order, log, one-hot, fixed-point encodings) \cite{teplukhin2019qae, mizuno2024optimal}, which can result in a much larger QUBO matrix than the problem’s natural definition. Some problems have inefficient, superlinear integer/float to binary encodings while others have efficient, linear or logarithmic scaling to the QUBO and less pre-processing. Also, various problems have higher-order terms to be mapped from a HOBO (higher order binary optimization) to QUBO problem, which can add superlinear scaling to the QUBO matrix size. For instance, Brubaker et al. \cite{brubaker2025quadratic} benchmarked QUBO optimization for the peptide-protein docking problem against constraint programming (CP), finding that mapping the problem to a HOBO binary optimization matrix was expensive in time, and also having to convert from HOBO to QUBO. This contributed to their conclusion that QUBO optimization is not a good fit for the peptide-protein docking problem.
\\
Post-processing methods to improve solutions are also important to evaluate. Often, QUBO solvers return infeasible or suboptimal solutions which can be improved upon.
For instance, steepest descent is a greedy algorithm to improve solutions, involving choosing a single bit flip at each iteration that minimizes the energy most and can be performed for a specified number of iterations. Problem-specific adjustments also exist to enforce feasible solutions or improve optimality, which may be simple normalization operations or adjustment of quantities that are represented by multiple bits in a QUBO (e.g. integers) \cite{teplukhin2019qae,mizuno2024optimal}. Evaluating the time complexity of such post-processing in addition to the actual solve time is important to determine the efficiency of a QUBO-based optimizer.

\subsection{QUBO analysis metrics}
Several metrics are important to evaluate the mathematical and structural aspects of QUBO formulations for optimization problems and are explained in the following sections.

\subsubsection{Quantitative connectivity}
One metric is the quantifiable connectivity between binary variables in the cost and penalty functions. Certain kinds of interconnectivity can make combinatorial optimization problems much harder to solve with classical computing methods as well as with quantum-inspired algorithms or via quantum device-based methods with limited interconnectivity. The quantum and quantum-inspired solvers considered here can handle arbitrary interconnectivity through methods including unique GPU-based annealing (Fujitsu DA) and hybrid quantum annealing (D-Wave HQA).
\\
Within quantifiable interconnectivity, the 3 following numerical metrics can be computed from a QUBO:
\begin{itemize}
  \item Size
  \item Density
  \item Interconnectivity
\end{itemize}
\textbf{Size} is the number of variables in a QUBO, which can make the problem more resource-heavy to solve due to the expansion of the number of possible solutions (combinatorial explosion). In the case of quantum annealing, larger sizes also require more qubits when each variable is represented by a qubit.
For a given variable vector $x \in \mathbb{R}^s$, the size of the QUBO is
\begin{equation}
  \text{Size} \equiv |x|
\end{equation}
\\
\textbf{Density} is defined as the ratio of nonzero QUBO terms to the total number of terms ($|x|^2$). Such terms are present in a cost or constraint equation $Q$ and in the format of $Q_{ij} x_ix_j$, with $Q_{ij} \neq 0$ and either $i = j$ or $i \neq j$. A QUBO with all coefficients being zero (all $Q_{ij} = 0$) would have a density of 0, and on the other extreme, a QUBO with all coefficients being nonzero (all $Q_{ij} \neq 0$) would have a density of 1. In the context of quantum annealing, density captures the number of couplings between qubits due to the need for embedding onto QPU hardware, and in gate-based quantum computing it represents the number of two-qubit gates needed for the quantum approximation optimization algorithm (QAOA) \cite{nublein2024reducing, maurizio2025quantum}.
\\
Formally, the density of the QUBO $Q$ with $x$ variables is \begin{equation}
  \text{Density} \equiv \frac{\sum_{i, j} \begin{cases} 1 & \text{if } Q_{ij} \neq 0 \\ 0 & \text{if } Q_{ij} = 0 \end{cases}}{|x|^2}
\end{equation}
QUBO problems that only have linear terms in the form $Q_{ii}x_i$ will naturally have a very low density given this definition.
\\
\textbf{Interconnectivity} is defined as the average coupling ratio of variables in the QUBO. The coupling ratio is the ratio of number of couplings a variable has to the maximum possible number of couplings ($|x|$). One difference between this measure and density is that we count any two variables $x_i$ and $x_j$ that appear in the same constraint equation as coupled even if not multiplied by each other, since their quantities are essentially coupled in terms of solution feasibility and this is not captured by the density metric.
The interconnectivity of the QUBO $Q$ with $x$ variables is formally defined as \begin{equation}
  \text{Interconnectivity} \equiv \frac{ \sum_{i=1}^{|x|} \sum_{j} \begin{cases} 1 & \text{if } (Q_{ij} \neq 0) \lor \left( \bigvee_{p=1}^{|C|} (i \in \mathcal{V}_p \land j \in \mathcal{V}_p) \right) \\ 0 & \text{otherwise} \end{cases}}{|x|^2}
\end{equation}
where $C$ is the set of constraints, and $\mathcal{V}_p$ is the set of variable indices that have nonzero coefficients in constraint $C_p$.
\\
Another factor to be considered is the rank-1 dominance, or how much of the QUBO's connectivity can be explained by rank one QUBO terms. A rank one QUBO equation is in the form \begin{equation}
  x^TQx = \sum_{i, j} s_i s_j x_i x_j = \left(\sum_i s_i x_i\right)^2
\end{equation}
where $s$ are coefficients in the symmetric matrix $Q$, and $x$ is the vector of binary variables. A QUBO optimization problem with such a rank one matrix is solvable in polynomial time  \cite{ferrez2004solving,punnen2022quadratic}. As such, problems with high density or high interconnectivity that have the majority of their terms coming from rank one QUBO equations (they may have other QUBOs as well) may be easier to solve with classical solvers due to the tractability of rank one QUBO optimization.

\subsubsection{Penalty structure}
Constrained QUBOs can include penalty terms, which are additional terms that must sum to zero for a solution to be considered feasible.
\\
Feasibility is enforced with different constraint types, with the main types of constraints for optimization problems being linear, quadratic, and one-hot constraints (or a combination of these).
\\
A linear constraint is expressed as a sum of variables multiplied by coefficients with a requirement that the sum must be less than or equal to a constant. This is formally defined as \begin{equation}
  \sum_{i=1}^{|x|} a_i x_i \leq c
\end{equation}
where $a_i$ is a coefficient for variable $x_i$, and $c$ is a constant \cite{katayama2024}.
\\
A quadratic constraint can contain products of two variables, and is formally defined as \begin{equation}
  \sum_{i=1}^{|x|} \sum_{j=1}^{|x|} a_{ij} x_i x_j \leq c
\end{equation}
where $a_{ij}$ is a coefficient for the product of variables $x_i$ and $x_j$, and $c$ is a constant; this can include linear terms where $i = j$.
\\
A one-hot constraint is a constraint that requires one and only one variable out of a certain subset of variables to be set to 1, and the rest to 0 \cite{katayama2024}. Such a constraint is formally defined as \begin{equation}
  \sum_{i=j}^{k} x_i = 1
\end{equation}
where $x_i$ is a variable in the subset of variables from $j$ to $k$.
\\
Constrained QUBO problems with quadratic and/or one-hot constraints can potentially be more difficult to solve efficiently with classical solvers.
Thus, quantum and quantum-inspired solvers could possibly provide performance benefits in these cases especially.
\\
Penalty separation, or the separation between penalty equation terms (feasibility) and cost function terms (optimality), can also be evaluated. If one intends on applying QUBO solvers that do not support constraints terms, penalty terms cannot be expressed in a separate constraint equation and must be manually embedded into the cost function with a scalar (Lagrange multiplier). Such is the case in previous studies applying quantum annealing to ground state energy calculations \cite{teplukhin2021computing, teplukhin2019qae}, mRNA codon optimization \cite{fox2021}, and reaction network pathway analysis \cite{mizuno2024optimal}. These problems may require penalty tuning, or an outer optimization loop needed to tune parameters that scale penalty terms in a cost function to help the optimizer balance between focusing on optimality or feasibility \cite{teplukhin2019qae, teplukhin2021computing, mizuno2024optimal}. Such a process requires numerous runs of an optimizer and can add significant computational time. In other cases outer optimization may not be needed \cite{fox2021}, however the penalty within the cost function can disturb other terms in the cost function and lead to suboptimal solutions.

\subsection{Solver performance metrics}

The above sections explained the metrics for evaluating QUBO formulations of optimization problems, and this section focuses on the metrics for evaluating the performance of the different solvers on such problems.
\\
Many solvers support penalty terms in a QUBO such as the Fujitsu DA, D-Wave HQA, and classical MIP and CP solvers (Gurobi, HiGHS, SCIP, CP-SAT).
When comparing to classical MIP or CP solvers one has to reformulate the QUBO problem.
Different metrics including accuracy of solution and the time taken to obtain a solution of a certain accuracy can be evaluated.
\\
For the D-Wave HQA solvers, different hybrid solver configurations are tested as highlighted in Table \ref{tab:dwave_hqa_configs}. The Leap Hybrid Nonlinear (NL) solver which supports decision-variable based problems and constraints, the Leap Hybrid Constrained Quadratic Model (CQM) solver which supports constraints, and the Leap Hybrid Binary Quadratic Model (BQM) solver are benchmarked \cite{dwave_leap_hybrid}. Additionally, solvers from the dwave-hybrid open source framework \cite{dwave_hybrid_framework} which support embedding constraints into the cost function with a Lagrange multiplier, are tested. The Leap solvers run both CPU and QPU calculations on D-Wave's cloud infrastructure, whereas for the dwave-hybrid framework the CPU calculations are performed locally. A pure QA solver is not benchmarked due to large problem sizes with heavy interconnectivity that are incompatible with QPU embeddings; such embedding incurs a quadratic overhead \cite{maurizio2025quantum}.
\begin{table}[H]
  \footnotesize
  \centering
  \begin{threeparttable}
    \caption{D-Wave HQA solver configurations}
    \label{tab:dwave_hqa_configs}
    \begin{tabularx}{\linewidth}{|>{\centering\arraybackslash}X|>{\centering\arraybackslash}X|>{\centering\arraybackslash}X|}
      \hline
      \textbf{Solver Configuration} & \textbf{Description} & \textbf{Framework} \\
      \hline
      Kerberos Hybrid & SA \& Tabu Search on CPU, & dwave-hybrid \\
      & QPU for high-impact subproblems &  \\
      \hline
      Hybrid Parallel Tempering (PT) & PT on CPU, QPU for high-impact subproblems  & dwave-hybrid \\
      \hline
      Leap Hybrid BQM & Out-of-the-box hybrid solver with classical heuristics and QPU & Cloud service \\
      \hline
      Leap Hybrid CQM & Out-of-the-box hybrid solver with classical heuristics and QPU, supports constraints & Cloud service \\
      \hline
      Leap Hybrid NL & Out-of-the-box hybrid solver with classical heuristics and QPU, supports decision variables & Cloud service \\
      \hline
    \end{tabularx}
  \end{threeparttable}
\end{table}
The Hybrid PT solver is a custom workflow developed with the dwave-hybrid framework \cite{dwave_hybrid_framework} that combines solutions from PT with solutions from high-impact subproblems of up to 50 variables solved on the QPU. To benchmark accuracy and efficiency, the following metrics are used:
\begin{itemize}
  \item The optimal value (minimum cost) obtained in a solution for a given problem.
  \item The time taken to obtain the minimum cost solution for a given problem (time to solution).
\end{itemize}
These metrics are calculated for each problem in a dataset and then averaged over the dataset to obtain the average cost (AC) and average time to solution (ATTS). AC is the average minimum objective value obtained over all problems, and ATTS is the average time to solution across all problems. AC and ATTS are calculated for each solver and then compared between the different solvers to analyze overall accuracy and efficiency; a plot of time to solution versus problem size is also created to gauge the scalability of the different solvers.

\section{Use cases benchmarking}
\label{sec:use_cases}

\subsection{Reaction network pathway analysis}

\begin{sidewaysfigure}[!htbp]
  \centering
  \includegraphics[width=1\linewidth]{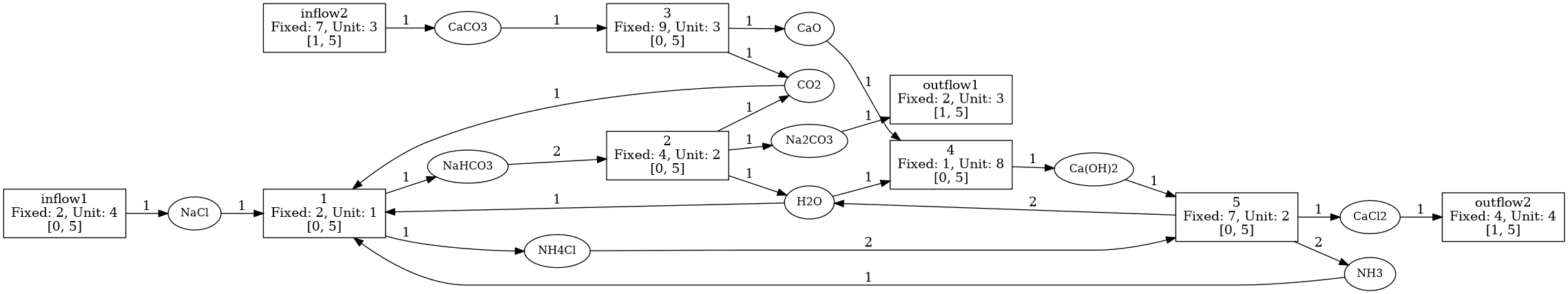}
  \caption{CRN graph for the Solvay process}
  \label{fig:solvay_crn}
\end{sidewaysfigure}

\begin{sidewaysfigure}[!htbp]
  \centering
  \includegraphics[width=1\linewidth]{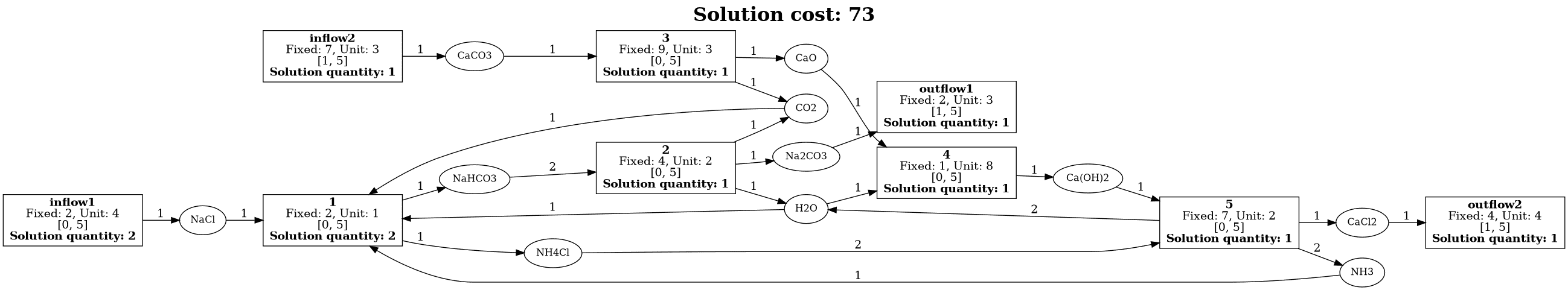}
  \caption{Solution with optimal reaction pathway for the Solvay process}
  \label{fig:solvay_crn_solution}
\end{sidewaysfigure}

Chemical reaction networks (CRNs) have numerous different pathways to produce a certain target material of interest, and solving this constrained combinatorial optimization problem has applications for synthesis planning and metabolic pathway analysis \cite{mizuno2024optimal}. Furthermore, finding the optimal pathway to maximize output of a target chemical is in general an NP-hard problem \cite{andersen2012} due to the combinatorial explosion from the network graph size that causes an exponential increase in runtime for a classical algorithm.
\\
A CRN can be represented as a graph with two kinds of nodes: reactions and species. This is a bipartite graph as all edges are between a species and a reaction, and a directed graph since an edge can either go from a species to a reaction (reactant) or from a reaction to a species (product) \cite{mizuno2024optimal}.
\\
We model each reaction as having a unit cost (cost per quantity of the reaction) to model a purchase price and a fixed cost (cost occurring if the reaction happens at all). A unit cost can represent costs to purchase substrates and dispose of byproducts, whereas a fixed cost can for instance model equipment preparation costs that do not vary with the number of times a reaction occurs \cite{mizuno2024optimal}. Both of these costs will vary depending on the reaction involved. A reaction also has a lower and upper bound for the number of times it can occur.
\\
Figure \ref{fig:solvay_crn} shows an example CRN for the Solvay process which is an industrial chemical process to form soda ash (Na\textsubscript{2}CO\textsubscript{3}). Squares represent reactions and circles are species, with each arrow representing the quantity of species to/from a reaction and each reaction having the $[l, u]$ meaning it can occur between $l$ and $u$ times.
\\
Figure \ref{fig:solvay_crn_solution} shows the optimal reaction pathway quantities in this network for producing soda ash (a simple network where all reactions are occurring).

\begin{figure}[H]
  \centering
  \includegraphics[width=.5\linewidth]{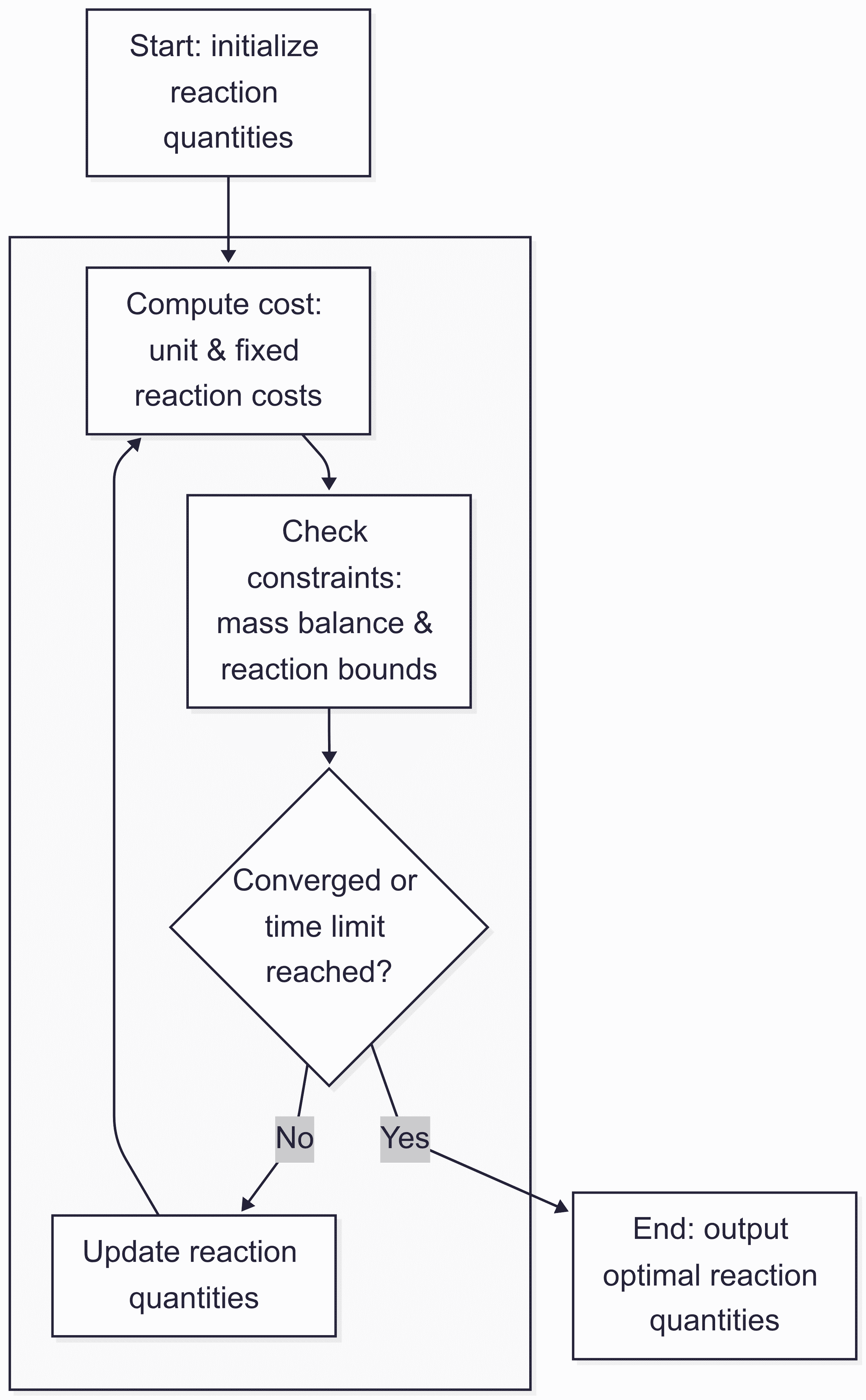}
  \caption{Generic CRN optimization flow}
  \label{fig:crn_flowchart}
\end{figure}

\newpage
\noindent
The overall combinatorial optimization problem for identifying the optimal reaction network pathway can be formulated as
\begin{align}
  \text{min } \sum_{r \in R} c_r^\text{unit}x_r + c_r^\text{fixed} p(x_r) \nonumber \\
  \text{s.t. } \forall s \in S, \sum_{r \in R} v_{s, r}x_r = 0, \nonumber \\
  \forall r \in R, l_r \leq x_r \leq u_r
\end{align}
where $R$ is the set of reactions and $S$ is the set of species, $c_r^\text{unit}$ denotes the unit cost of a reaction and $c_r^\text{fixed}$ the fixed cost, $x_r$ the quantity of a reaction (the variables to optimize), $v_{s, r}$ the signed stoichiometric coefficient of a species in a reaction, and $l_r$ and $u_r$ the bounds for a reaction quantity \cite{mizuno2024optimal}.
\\
The positivity indicator function $p(x_r)$ is used to encode whether a reaction should incur a fixed cost due to it having a positive quantity, and is defined as:
\begin{equation}
  p(x_r) =
  \begin{cases}
    1 & \text{if } x_r > 0 \\
    0 & \text{if } x_r = 0
  \end{cases}
\end{equation}
\\
To facilitate the positivity indicator in a MILP, CP, or QUBO context, one can define $p(x_r) = y_r$ where $y_r$ is a binary variable in the optimization problem with the constraint
\begin{equation}
  x_r \leq u_r y_r
\end{equation}
to enforce that $y_r$ is set correctly based on $x_r$.
\\
This problem flow for finding the optimal reaction network pathway is summarized in Figure \ref{fig:crn_flowchart}.

\noindent
The stoichiometric equation constraint for each species (mass balance equation) can be encoded as a constraint in the problem formulation for a MILP/CP problem. For the QUBO format of the Fujitsu DA, one has to specify two inequality constraints for the mass balance of each species, $s$, because multiple equality constraints are not supported:
\begin{align}
  \sum_{r \in R} v_{s, r}x_r \geq 0 \nonumber \\
  \sum_{r \in R} v_{s, r}x_r \leq 0
\end{align}
These constraint equations being separate from the cost function prevents one from having to do a manual penalty-strength tuning loop and run several annealings per problem. That process of penalty-strength tuning led to the need for post-processing, and increased runtime in the work which originally solved this problem with annealing methods \cite{mizuno2024optimal}.
\\
Moreover, the unary and log integer encoding methods for the Fujitsu DA QUBOs were tested for encoding the $x_r$ integers into bits, as they were found to be the most accurate and efficient (least number of bits) methods, respectively \cite{mizuno2024optimal}.
\\
To express an integer $x \in [l, u]$, one can use $n$ bits $q \in \{0, 1\}^n$. Unary encoding requires $d = u - l$ bits and encodes the integer as \begin{equation}
    l + \sum_{k = 1}^{d} q_k
\end{equation} Log encoding requires $K + 1$ bits where $K = \lfloor \log_2 d \rfloor$, and the integer is encoded as \begin{equation}
  l + \left(\sum_{k = 1}^{K - 1} 2^k q_k\right) + \left(d - (2^K - 1)\right) q_K
\end{equation}

\begin{figure}[H]
  \centering
  \includegraphics[width=1\linewidth]{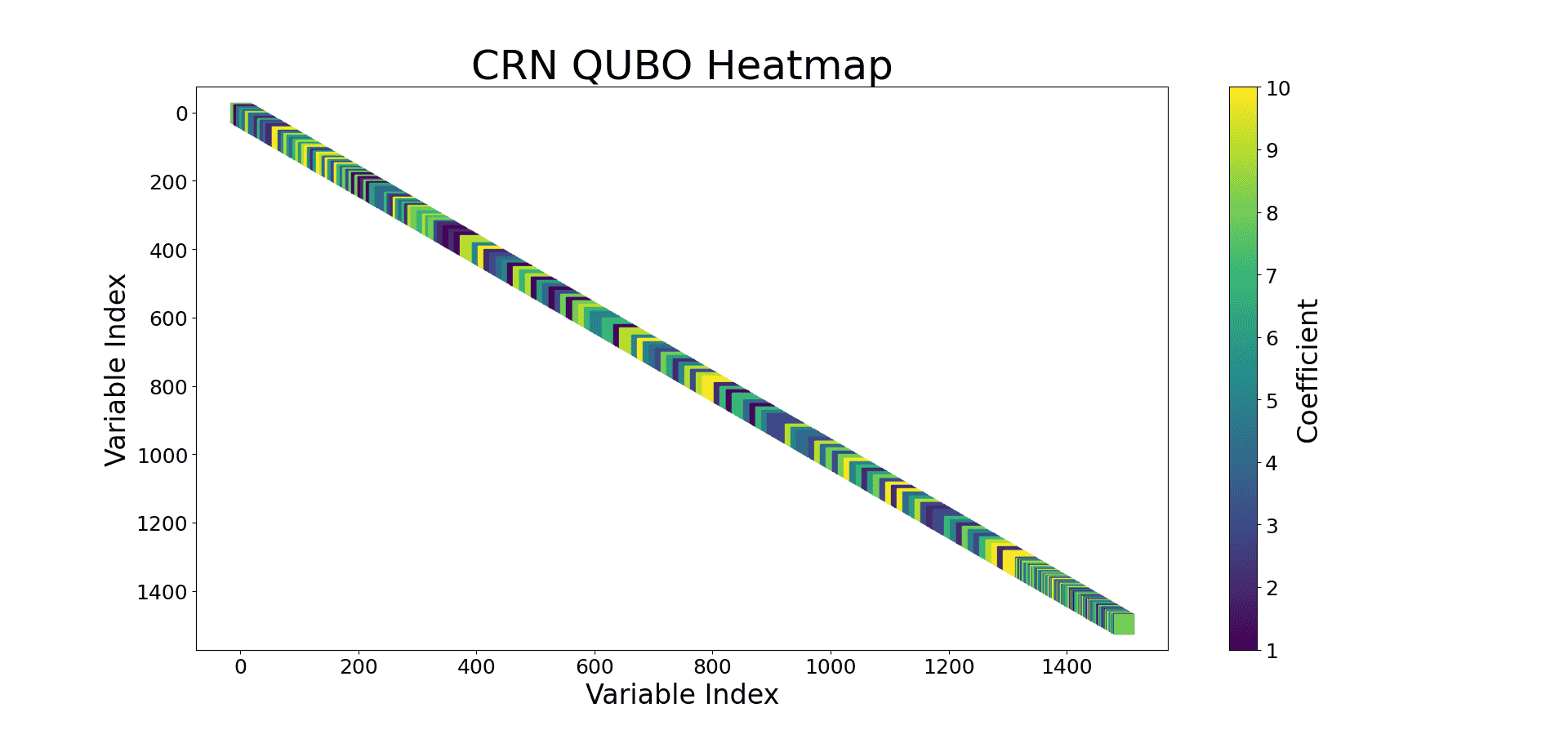}
  \caption{Heatmap of QUBO structure from USPTO CRN dataset}
  \label{fig:uspto_small_heatmap}
\end{figure}

\noindent
A dataset was formed by selecting reaction networks from the USPTO patent chemical reactions datasets \cite{lowe2017patents}, similar to what was done by Mizuno et al. \cite{mizuno2024optimal}; here 100 species or less per network were included.
\\
A heatmap for the coefficients in a cost function QUBO for a reaction network in the USPTO dataset is shown in Figure \ref{fig:uspto_small_heatmap}. One can see the low density in this QUBO, as it does not have quadratic terms given the linear cost function. The coefficients in the cost function vary from 1 to 10 which were the minimum and maximum cost values included in these optimization problems respectively. This problem has no rank-1 dominance as it only contains linear terms.
\\

\begin{sidewaysfigure}[!htbp]
    \centering
    \includegraphics[width=1\linewidth]{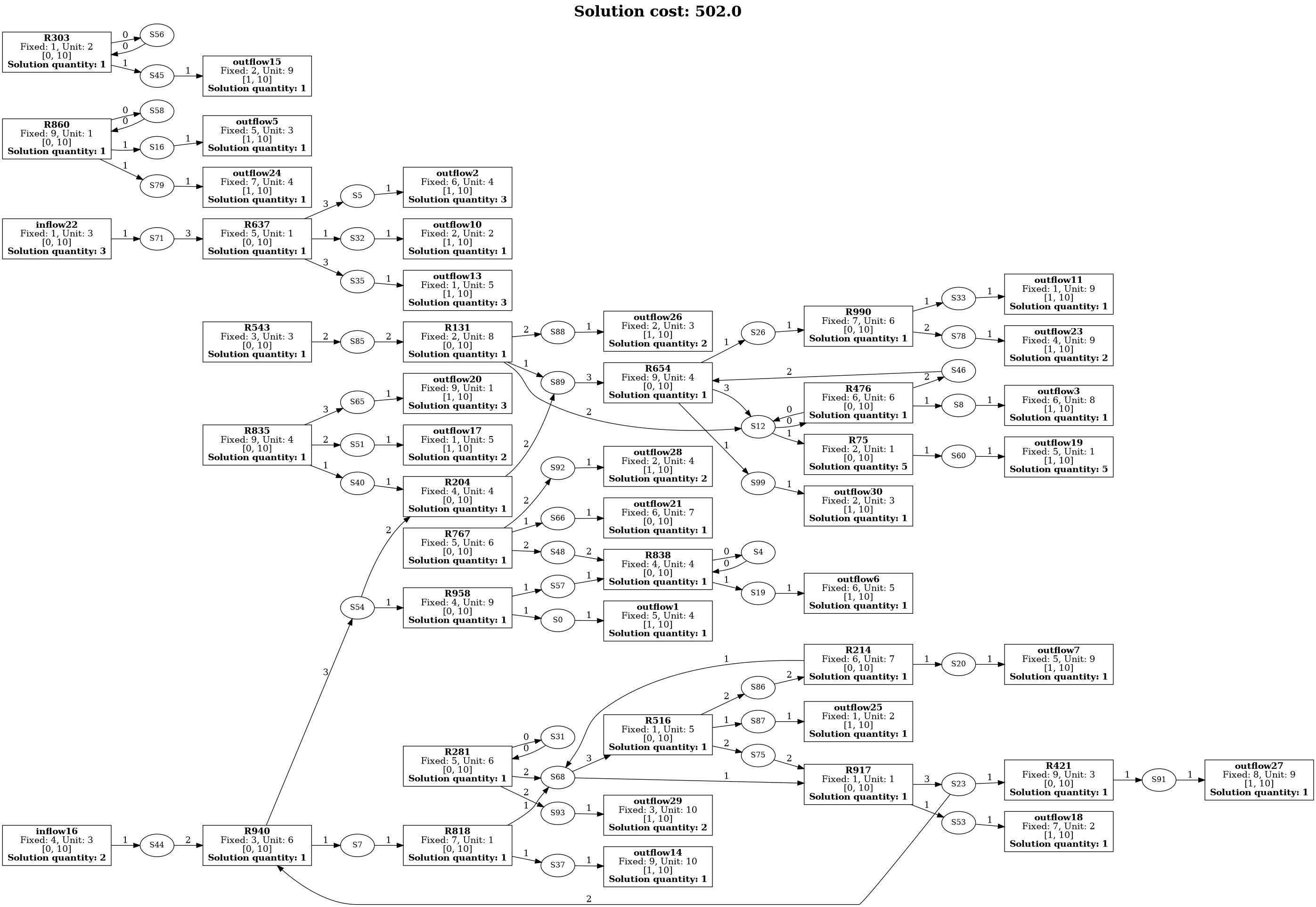}
    \caption{Artificial CRN solution (small subgraph of full CRN)}
    \label{fig:artificial_crn}
\end{sidewaysfigure}

\noindent
Furthermore, an artificial dataset of 3 made-up reactions with high interconnectivity between species and reactions (100 species, 10 random reactions per species) was created to make a dataset for which classical heuristic solvers may potentially struggle.
\\
While these networks are too large to visualize fully, Figure \ref{fig:artificial_crn} shows the reaction quantities and species that take part in the optimal reaction pathway (excluding all other species and reactions).

\clearpage
\newpage
\noindent
Tables \ref{qubo_metrics}, \ref{solver_comparison_uspto}, \ref{solver_comparison_artificial} show the QUBO metrics and solver comparisons from the current work on all datasets. The Digital Annealer v4 was used, and for classical MIP/CP solvers Gurobi version 12.0.1, HiGHS version 1.11.0, CP-SAT version 9.14.6206, and SCIP version 9.0 were run on a server with 48 cores and 98 GB of RAM.

\begin{table}[H]
  \scriptsize
  \centering
  \begin{threeparttable}
    \caption{CRN QUBO comparison}
    \label{qubo_metrics}
    \begin{tabularx}{\linewidth}{@{}>{\centering\arraybackslash}X>{\centering\arraybackslash}X>{\centering\arraybackslash}X>{\centering\arraybackslash}X>{\centering\arraybackslash}X@{}}
      \toprule
                                           & \textbf{USPTO} &             & \textbf{Artificial} &             \\
      \midrule
      \textbf{Dataset Size}              & 12               &             & 3                           &             \\
      \textbf{Encoding}                  & Unary            & Log         & Unary                       & Log         \\
      \textbf{Encoding Scaling}          & Linear           & Logarithmic & Linear                      & Logarithmic \\
      \textbf{Avg. Size}              & 975.75           & 512.083     & 11540.133                   & 5254        \\
      \textbf{Avg. Density}           & 0.001            & 0.002       & 8.67e-5                     & 1.90e-4     \\
      \textbf{Avg. Int.} & 0.05             & 0.055       & 0.106                       & 0.082       \\
      \bottomrule
    \end{tabularx}
  \end{threeparttable}
\end{table}

\begin{table}[H]
  \centering
    \caption{Solver comparison for USPTO dataset}
    \label{solver_comparison_uspto}
    \begin{tabularx}{\textwidth}{XXXX}
      \toprule
      \textbf{Name} & \textbf{Encoding} & \textbf{AC} & \textbf{ATTS*} \\
      \midrule
      CP-SAT        & N/A      & 132.417  & <0.02    \\
      Gurobi        & N/A      & 132.417  & <0.02    \\
      HiGHS         & N/A      & 132.417  & <0.02    \\
      SCIP          & N/A      & 132.417  & <0.02    \\
      DA            & Log      & 132.500  & 30.615   \\
      DA            & Unary    & 135.250  & 20.610   \\
      \bottomrule
    \end{tabularx}
    \caption*{
    \textbf{AC}: average cost; average final solution cost over all problems \\
    \textbf{ATTS [s]}: average time to solution in seconds \\
    * Columns with ATTS on the order of milliseconds are not diﬀerentiated as such minute time diﬀerences can be due to external factors such as server load}
\end{table}

\begin{table}[H]
  \centering
    \caption{Solver comparison for artificial dataset}
    \label{solver_comparison_artificial}
    \begin{tabularx}{\textwidth}{XXXX}
      \toprule
      \textbf{Name} & \textbf{Encoding} & \textbf{AC} & \textbf{ATTS} \\
      \midrule
      Gurobi   & N/A      & 455.667  & 3528.728 \\
      CP-SAT   & N/A      & 455.667  & 3530.501 \\
      HiGHS    & N/A      & 460.333  & 3530.501 \\
      SCIP     & N/A      & 1170.667 & 3530.005 \\
      DA       & Unary    & 1842.000 & 2000.959 \\
      \bottomrule
    \end{tabularx}
    \caption*{
    \textbf{AC}: average cost; average final solution cost over all problems \\
    \textbf{ATTS [s]}: average time to solution in seconds}
\end{table}
\noindent
While the classical Gurobi and CP-SAT solvers struggle in runtime on the artificial dataset problems (as shown in Table \ref{solver_comparison_artificial}) when solved to optimality (minimum value of the cost function), the Fujitsu DA is not able to find better or near-optimal solutions; the classical solvers were time-limited for the artificial dataset, and HiGHS and SCIP did not obtain optimal solutions. Even when the digital annealer was started with a near-optimal solution from Gurobi (run for 100 seconds), it was not able to identify any solutions better than the initial given solution from the classical solver. Overall, the classical MIP/CP solvers are able to solve the CRN problem to optimality, while the digital annealer has significantly higher solve times on the smaller USPTO dataset and it is unable to identify near-optimal solutions for the artificial dataset. These results are in agreement with the work by Mizuno et al. \cite{mizuno2024optimal} which found that the runtime of Gurobi to solve this problem scaled more efficiently than the runtime with QA (non-hybrid) or SA. The chemical reaction network pathway analysis problem which has low density, a linear cost function, and linear constraints, does not appear to have utility from quantum and quantum-inspired solvers based on these results and those of Mizuno et al. \cite{mizuno2024optimal}.

\subsection{mRNA codon selection}

\begin{figure}[H]
  \centering
  \includegraphics[width=0.8\linewidth]{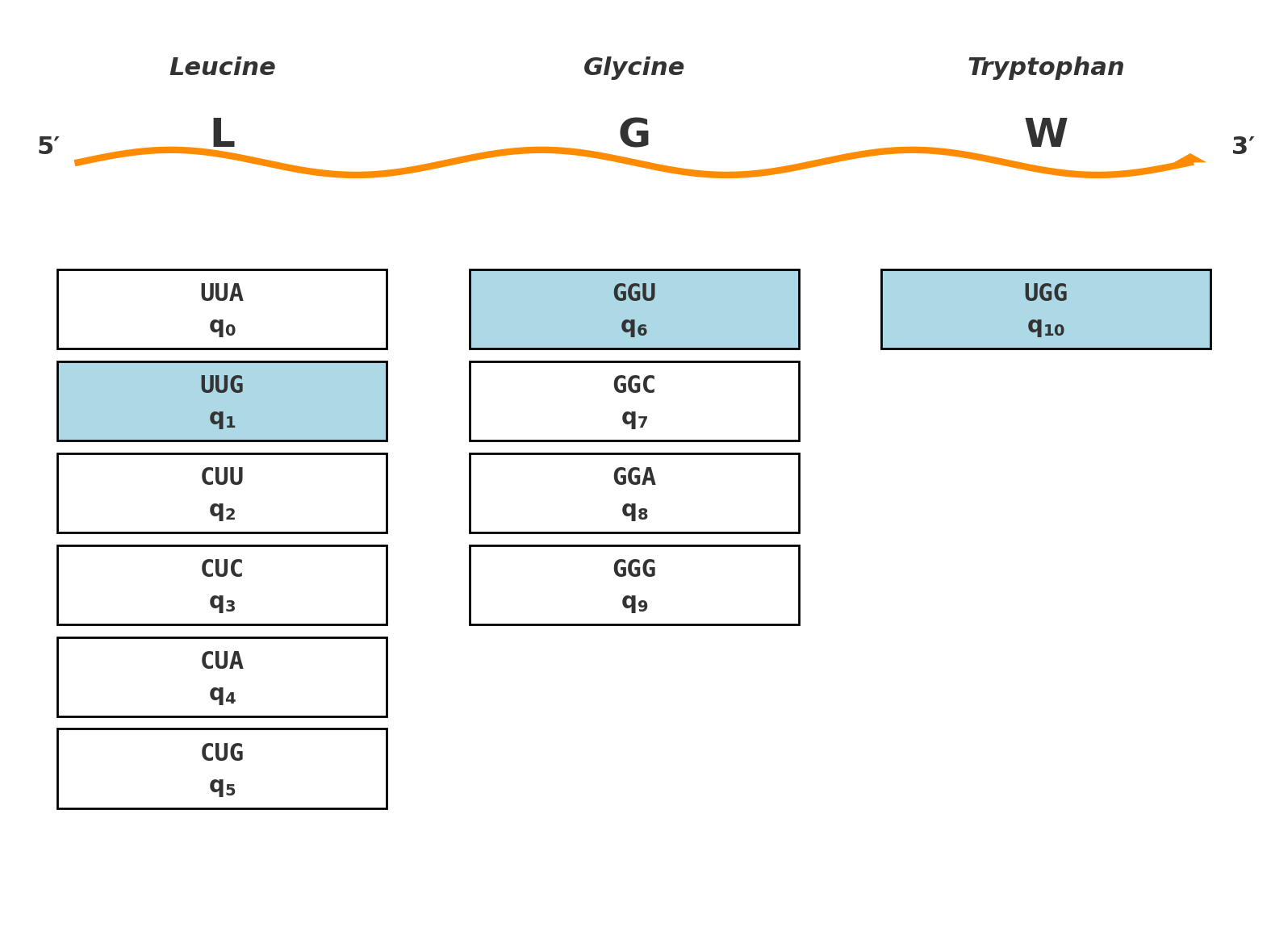}
  \caption{Possible mRNA codon choices}
  \label{fig:codon_choices}
\end{figure}

\begin{figure}[H]
  \centering
  \includegraphics[width=1\linewidth]{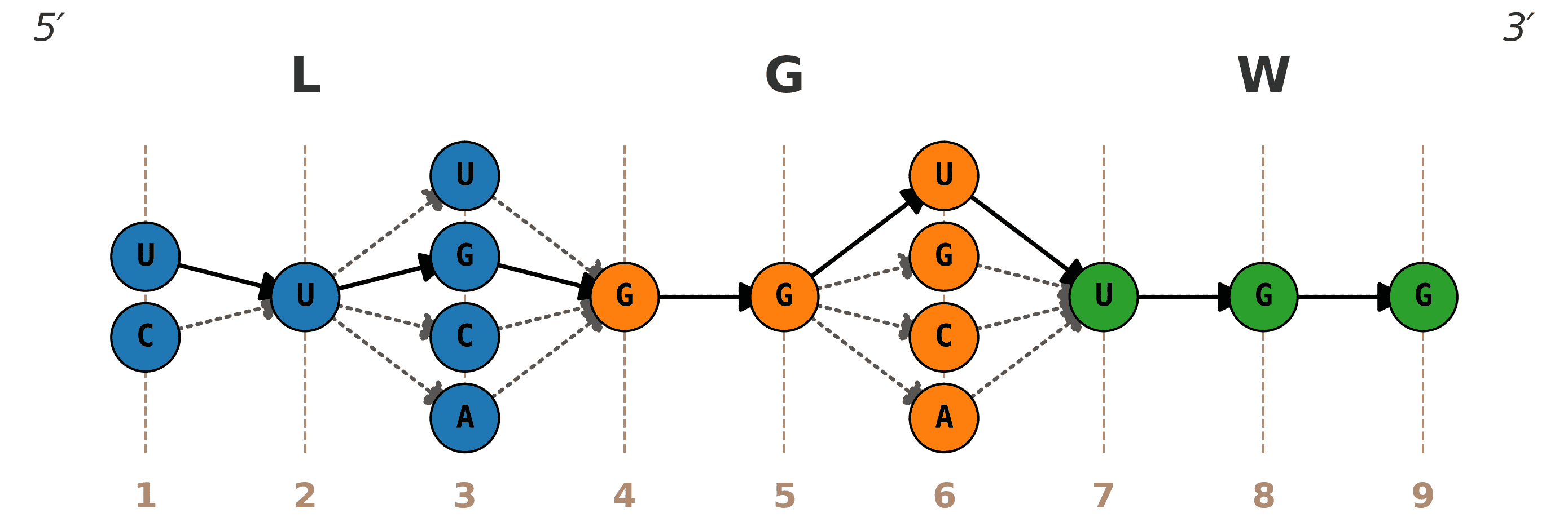}
  \caption{Graph representation of possible nucleotide choices}
  \label{fig:codon_choices_graph}
\end{figure}

The translation of protein sequences into efficient Messenger RNA (mRNA) represents a complex NP-hard combinatorial problem. mRNA optimization addresses the critical challenge of enhancing gene expression levels by modifying codon sequences. The fundamental goal of codon optimization is to select codons that maintain a certain amino acid sequence to encode a protein, but maximize the expression probability in a given host organism; this expression depends on various stochastic biochemical reactions and is extremely difficult to compute directly \cite{salis2009automated}. Achieving this involves navigating complex trade-offs between several competing biological factors, including codon bias, content of G and C nucleotides, mRNA secondary structure, mRNA folding stability around the ribosome, and more \cite{fox2021, kim1997, brule2017, kudla2009coding}. Codon choice has been shown to affect protein folding and functions such as channeling \cite{buhr2016synonymous, kirchner2017alteration} which are useful in different applications including recombinant protein drugs and nucleic acid therapeutics \cite{mauro2014critical}.
\\
The genetic code's degeneracy causes the optimization complexity. For instance, leucine, a common amino acid, can be encoded by six distinct codons: CUA, CUC, CUG, CUU, UUA, and UUG. Each of these codons may exhibit significantly different expression efficiencies depending on the host organism. For a protein sequence of just 100 amino acids, with each position having an average of 3 synonymous codon options, the solution space encompasses approximately $3^{100}$ ($5 * 10^{47}$) possible nucleotide combinations, while biologically-relevant sequences can have thousands of amino acids.
\\
Specifically, optimizing mRNA sequences necessitates formulating these biological constraints as a combinatorial optimization problem, which can be computationally challenging due to the exponential growth of possible codon combinations.
\\
This work, along with the work by Fox et al. \cite{fox2021}, does not evaluate in which ways to perform codon optimization for biological applications, but rather focuses on the feasibility of using quantum and quantum-inspired solvers to solve this problem with its NP-hard complexity.
\\
For genomics optimization problems in general a risk is mapping overhead to QUBO; converting constraints from a problem into a QUBO formulation can incur significant overhead in memory and time. As presented by Maurizio and Mazzola, such overhead is especially common when constraints restrict a large subset of the solution space, very common in genomics problems such as mRNA codon selection \cite{maurizio2025quantum}; constrained QUBO formulations and solvers can be leveraged to reduce this overhead.
\\
The overall combinatorial optimization problem for codon selection is expressed as a Quadratic Unconstrained Binary Optimization (QUBO) in which every possible codon for each amino acid is
represented by a binary variable \(q_i\in\{0,1\}\).  A value of \(1\) indicates that codon \(i\) is chosen for the respective amino acid, while \(0\) means it is ignored.  For a protein with \(L\) amino acids, the decision space contains
\begin{equation}
  |q| = \sum_{k=1}^{L} n_k
\end{equation}
variables, where \(n_k\) is the number of codons that can encode the \(k^{\text{th}}\) amino-acid. This representation of possible codon choices for each amino acid is shown in Figure \ref{fig:codon_choices}, containing one possible feasible selection; a graph representation of the different nucleotide choices for each codon is shown in Figure \ref{fig:codon_choices_graph}, inspired by the visualization in \cite{ward2025mrna}. The corresponding binary variables $q$ would be represented as

\begin{equation}
  q = \left[
    \begin{array}{ccccccccccc}
      0 & 1 & 0 & 0 & 0 & 0 & 1 & 0 & 0 & 0 & 1
    \end{array}
  \right]
\end{equation}
and the amino acids in the sequence involve leucine ($n_1 = 6$), glycine ($n_2 = 4$), and tryptophan ($n_3 = 1$).

\noindent
This problem can be described with a cost function (Hamiltonian) which will be described further in the following paragraphs, of the QUBO matrix form
\begin{equation}
  H \;=\; \sum_i Q_{ii} q_i + \sum_{i<j} Q_{ij} q_i q_j \;=\; H_f \;+\; H_{GC} \;+\; H_{R}
\end{equation}
with each of the three terms capturing a distinct biological preference. The Hamiltonian does not include embedded constraints in this work which applies constraint-based solvers, hence separate one-hot constraints $P$ are used to enforce feasibility which is described below.

\paragraph{Codon usage bias (\(H_f\)).}
Given that codon usage differs among different kinds of organisms \cite{grantham1980codon}, it is important to favor codons that are abundant in the expression host.
To encode this preference, a contribution
\begin{equation}
  H_f \;=\; c_f \sum_{i=1}^N \Bigl[\!\log\!\Bigl(\frac{1}{C_i}+\varepsilon_f\Bigr)\Bigr]\,q_i
\end{equation}
is introduced, where \(N = |q|\) is the total number of possible codons, \(C_i\) is the usage frequency of codon \(i\) from 0 (exclusive) to 1 (inclusive), \(c_f>0\) is a scaling coefficient and \(\varepsilon_f = 1\) is an offset for the log function used for consistency with the work of Fox et al. \cite{fox2021} even though it is not required since $1/C_i \geq 1$ and $C_i \neq 0$.  Rare codons therefore incur large positive penalties, while common codons contribute less.

\paragraph{GC content control (\(H_{GC}\)).}
Organisms have varying preferences for the GC content of their mRNA sequences. Maintaining the global fraction of \(\mathrm{G}+\mathrm{C}\) bases, \(\rho_{GC}\), close to a desired target
\(\rho_T\) specific to the host organism, is enforced through quadratic terms
\begin{align}
  H_{GC} &\;=\; c_{GC}(\rho_{GC} - \rho_T)^2 \nonumber \\
  &\;=\; \frac{2c_{\mathrm{GC}}}{N^{2}}
  \sum_{i=0}^{N - 1}\sum_{j<i}^{N - 1} s_i s_j\,q_i q_j
  + \frac{c_{\mathrm{GC}}}{N^{2}}
  \sum_{i=0}^{N - 1} s_i^{2}\,q_i
  - \frac{2\rho_T c_{\mathrm{GC}}}{N}
  \sum_{i=0}^{N - 1} s_i\,q_i
  + c_{\mathrm{GC}}\rho_T^{2}
\end{align}
where \(s_i\in\{0,1,2,3\}\) counts the number of \(\mathrm{G}\) and \(\mathrm{C}\) nucleotides in codon
\(i\) and \(c_{GC}\) weights the importance of this property. The sum includes upper triangular terms in the QUBO matrix multiplied by 2, as the quadratic terms are symmetric \cite{fox2021}.\\
These GC terms in the Hamiltonian add significant connectivity to the QUBO due to the quadratic couplings between all codons of different amino acids. However, the quadratic terms $H_{GC}$ form a rank one QUBO equation, which as explained in Section \ref{sec:qubo_metrics} can be solved in polynomial time by classical solvers; this does not imply the entire mRNA codon selection problem is solvable in polynomial time, as it is NP-hard.

\paragraph{Repeated nucleotide minimization (\(H_{R}\)).}
The cost function also includes terms to minimize the number of repeated nucleotides in the mRNA sequence. For every pair of codons
\((i,j)\) that occupy adjacent amino-acid positions a two-body cost
\(R_{ij}\,q_i q_j\) is added, where \(R_{ij}\) is the maximum number of repeated nucleotides squared in the two sequential codons, minus one since there will always be one repeated nucleotide. For example, for the codons ATA and TCG (ATATCG), \(R_{ij} = 1^2 - 1 = 0\) since the two codons have a maximum of 1 repeated nucleotide, while for the codons CGG and GGG (CGGGGG), \(R_{ij} = 5^2 - 1 = 24\). Summing over all pairs of codons gives
\begin{equation}
  H_{R} \;=\; c_R \sum_{i=0}^{N - 1}\sum_{j<i}^{N - 1} R_{ij} \kappa_{ij} \,q_i q_j
\end{equation}
with \(c_R\) a tunable weight and \(\kappa_{ij}\) a binary variable that is 1 if the two codons are in adjacent amino acids and 0 otherwise \cite{fox2021}. In practice this can be represented as a sparse matrix, only filling out terms for codons in adjacent amino acids.

\paragraph{One-hot constraints (\(P\)).}
While the work in \cite{fox2021} uses additional terms in the cost function to enforce the constraint that exactly one codon is selected for each amino acid, the current work uses a one-hot constraint for this purpose to strictly enforce the constraint and avoid the need for tuning a penalty weight for the constraint introduced in the cost function.
\\
The one-hot constraint enforces feasibility, ensuring for each amino acid with possible codon choices $\{q_i, \dots, q_j\}$ exactly one of the codons is selected, i.e.
\begin{equation}
\sum_{k = i}^{j} q_k = 1.
\end{equation}
The overall optimization problem as given by $H$ thus becomes 
\begin{align}
  \text{min } & c_f \sum_{i=1}^N \Bigl[\!\log\!\Bigl(\frac{1}{C_i}+\varepsilon_f\Bigr)\Bigr]\,q_i \nonumber \\
             & + \frac{2c_{\mathrm{GC}}}{N^{2}}
             \sum_{i=0}^{N - 1}\sum_{j<i}^{N - 1} s_i s_j\,q_i q_j
             + \frac{c_{\mathrm{GC}}}{N^{2}}
             \sum_{i=0}^{N - 1} s_i^{2}\,q_i
             - \frac{2\rho_T c_{\mathrm{GC}}}{N}
             \sum_{i=0}^{N - 1} s_i\,q_i
             + c_{\mathrm{GC}}\rho_T^{2} \nonumber \\
             & +  c_R \sum_{i=0}^{N - 1}\sum_{j<i}^{N - 1} R_{ij} \kappa_{ij} \,q_i q_j \nonumber \\
  \text{s.t. } & \sum_{k = i}^{j} q_k = 1, \nonumber \\
             & \forall \text{ amino acid positions } \{q_i, \dots, q_j\}
\end{align}

\paragraph{Formulation for non-constraint-based QUBO solvers.}
Constraint-based QUBO solvers including the Fujitsu DA and D-Wave Leap CQM HQA natively support constraints as a separate entity from the cost function. On the other hand, non-constraint-based QUBO solvers such as D-Wave's QA, Leap BQM HQA, and HQA solvers developed with the dwave-hybrid framework do not natively support constraints, requiring one to embed the constraint in the cost function. For such solvers, rather than including separate one-hot constraints, the constraints are embedded in the cost function Hamiltonian as a penalty term $H_P$ with the overall Hamiltonian then being
\begin{equation}
  H \;=\; H_f \;+\; H_{GC} \;+\; H_{R} \;+\; H_P
\end{equation}
Without constraints, a solution selecting zero codons for any amino acid would have the lowest cost according to the Hamiltonian. In the original work by Fox et al. \cite{fox2021} that applied QUBO solvers to mRNA codon optimization, $H_P$ was constructed as 
\begin{equation}
  H_{P} \;=\; -\epsilon\sum_{i=0}^{N - 1} q_i + \sum_{i=0}^{N - 1}\sum_{j<i}^{N - 1} \tau_{ij} \,q_i q_j
\end{equation}
where $\epsilon$ is a constant factor subtracted from each linear term $Q_{ii}$ in the Hamiltonian so that selecting zero codons is no longer the lowest-cost solution, and $\tau_{ij}$ is a large constant for codons $i, j$ of the same amino acid position (and 0 for other pairs of codons) to ensure that only one codon is selected for each amino acid. Epsilon was set such that
\begin{equation}
\epsilon > \max_{0 \le i \le N - 1} |Q_{ii}|
\end{equation}
and a value of
\begin{equation}  
  \tau_{ij\text{, same amino acid}} = 50 \, \max_{0 \le i \le N - 1} |Q_{ii}|
\end{equation}
was used for codons $i, j$ of the same amino acid position. A large enough value must be chosen to enforce the constraint, but too large of a value can affect numerical stability of the solver such as programmable field resolution of quantum annealers.\\
\\
This construction of $H_P$ developed by Fox et al. \cite{fox2021} was tested with unconstrained D-Wave HQA solvers and in the experiments done in this work, was unable to identify optimal solutions even for a toy problem of 3 amino acids; the QUBO formulation was verified with a D-Wave exact solver to ensure it was correct. Hence, this work proposes a direct one-hot constraint penalty term (standard for QUBO formulation) for $H_P$ instead:

\begin{equation}
  H_{P} \;=\; \mu \sum_{\{q_i,\dots,q_j\}} \left((\sum_{k = i}^{j} q_k) - 1\right)^2
\end{equation}
\noindent where the outer sum runs over all amino acid positions with codons $\{q_i, \dots, q_j\}$ and $\mu > 0$ weights the constraint strength; this penalizes solutions that do not select exactly one codon for each amino acid position. The strength scalar is set as
\begin{equation}
  \mu \;=\; 25\, \max_{0 \le i \le j \le N - 1} |Q_{ij}|
\end{equation}
with outer optimization over this Lagrange multiplier $\mu$ not found to be beneficial for solutions.
\\
Hence, the overall optimization problem with embedded constraints given by $H'$ thus becomes 
\begin{align}
  \text{min } & c_f \sum_{i=1}^N \Bigl[\!\log\!\Bigl(\frac{1}{C_i}+\varepsilon_f\Bigr)\Bigr]\,q_i \nonumber \\
             & + \frac{2c_{\mathrm{GC}}}{N^{2}}
             \sum_{i=0}^{N - 1}\sum_{j<i}^{N - 1} s_i s_j\,q_i q_j
             + \frac{c_{\mathrm{GC}}}{N^{2}}
             \sum_{i=0}^{N - 1} s_i^{2}\,q_i
             - \frac{2\rho_T c_{\mathrm{GC}}}{N}
             \sum_{i=0}^{N - 1} s_i\,q_i
             + c_{\mathrm{GC}}\rho_T^{2} \nonumber \\
             & +  c_R \sum_{i=0}^{N - 1}\sum_{j<i}^{N - 1} R_{ij} \kappa_{ij} \,q_i q_j \nonumber \\
             & + \mu \sum_{\{q_i,\dots,q_j\}} \left((\sum_{k = i}^{j} q_k) - 1\right)^2
\end{align}

\paragraph{Formulation for MIP solvers.}
For mixed-integer programming (MIP) solvers including HiGHS, Gurobi, and SCIP, the problem formulation can be linearized to eliminate density in the problem that was present in the QUBO due to the quadratic terms for the GC content optimization; those GC content quadratic terms represented a rank-1 matrix within the QUBO formulation. The QUBO formulation contained variables $q_i$ for a certain choice of a codon across all amino acids, requiring quadratic terms for codon interactions in the form of $q_i q_j$. On the other hand, the MIP formulation contains variables $x_{p,i}$ to represent selecting the $i$-th possible codon at amino acid position $p$ (as there are multiple choices due to the code's degeneracy), and variables $z_{p,i,j}$ to represent the transition from codon $i$ at position $p$ to codon $j$ at position $p+1$. This structure fits with the problem where we select one codon per amino acid position and need to account for nucleotide-level interactions between adjacent selected codons such as repeated nucleotides. The GC content calculation becomes linear since it can directly sum the number of GC nucleotides $s_{p,i}$ of selected codons without requiring quadratic terms, and the repetition terms $R_{i,j}$ between adjacent codons are stored per transition variables $z_{p,i,j}$ rather than through quadratic products. While the D-Wave Leap CQM HQA solver supports MIP formulations, this MIP formulation was not beneficial in its performance, whereas classical MIP/CP solvers did benefit from it.

\begin{align}
  H_{\text{MIP}} &= c_f \sum_{p=0}^{L-1} \sum_{i=0}^{m_p-1} \log\left(\frac{1}{C_{p,i}}+\varepsilon_f\right) x_{p,i} \nonumber \\
  &\quad + c_{GC} \sum_{k=0}^{3L}  \left(\frac{k}{3L} - \rho_T\right)^2 g_k \nonumber \\
  &\quad + c_R \sum_{p=0}^{L-2} \sum_{i=0}^{m_p-1} \sum_{j=0}^{m_{p+1}-1} R_{i,j} z_{p,i,j} \nonumber \\
\end{align}
\noindent where $x_{p,i} \in \{0,1\}$ indicates selection of codon $i$ at position $p$, $z_{p,i,j} \in \{0,1\}$ indicates transition from codon $i$ to codon $j$ between adjacent positions, $g_k \in \{0,1\}$ indicates whether the total GC count equals $k$, and $m_p$ is the number of possible codons for the amino acid at position $p$.
\\
Hence, the overall MIP optimization problem becomes 
\begin{align}
  \text{min } & c_f \sum_{p=0}^{L-1} \sum_{i=0}^{m_p-1} \log\left(\frac{1}{C_{p,i}}+\varepsilon_f\right) x_{p,i} \nonumber \\
  &\quad + c_{GC} \sum_{k=0}^{3L}  \left(\frac{k}{3L} - \rho_T\right)^2 g_k \nonumber \\
  &\quad + c_R \sum_{p=0}^{L-2} \sum_{i=0}^{m_p-1} \sum_{j=0}^{m_{p+1}-1} R_{i,j} z_{p,i,j} \nonumber \\
  \text{s.t. } & \sum_{i=0}^{m_p-1} x_{p,i} = 1, \quad \forall p \in \{0, \ldots, L-1\} \nonumber \\
             & \sum_{k=0}^{3L} g_k = 1 \nonumber \\
             & \sum_{p=0}^{L-1} \sum_{i=0}^{m_p-1} s_{p,i} x_{p,i} = \sum_{k=0}^{3L} k \cdot g_k \nonumber \\
             & \sum_{i=0}^{m_p-1} \sum_{j=0}^{m_{p+1}-1} z_{p,i,j} = 1, \quad \forall p \in \{0, \ldots, L-2\} \nonumber \\
             & z_{p,i,j} \leq x_{p,i}, \quad \forall p,i,j \nonumber \\
             & z_{p,i,j} \leq x_{p+1,j}, \quad \forall p,i,j
\end{align}

\paragraph{Formulation for constraint programming solvers.}
For pure constraint programming (CP) solvers such as CP-SAT, and the D-Wave Leap NL HQA solver which supports decision variables, the formulation uses integer decision variables $y_p \in \{0, \ldots, m_p-1\}$ that directly represent the chosen codon index at amino acid position $p$. Rather than using binary variables for each possible codon, $y_p$ serves as an index that selects the appropriate cost values. The CP formulation is given by
\begin{equation}
  H_{\text{CP}} = \sum_{p=0}^{L-1} c_f \log\left(\frac{1}{C_{p,y_p}}+\varepsilon_f\right) + c_{GC} \sum_{k=0}^{3L} \left(\frac{k}{3L} - \rho_T\right)^2 h_k + c_R \sum_{p=0}^{L-2} R_{y_p,y_{p+1}}
\end{equation}  
\noindent where $h_k \in \{0,1\}$ indicates whether the total GC count equals $k$.
\\
Hence, the overall CP optimization problem becomes 
\begin{align}
  \text{min } & \sum_{p=0}^{L-1} c_f \log\left(\frac{1}{C_{p,y_p}}+\varepsilon_f\right) + c_{GC} \sum_{k=0}^{3L} \left(\frac{k}{3L} - \rho_T\right)^2 h_k + c_R \sum_{p=0}^{L-2} R_{y_p,y_{p+1}}\nonumber \\
  \text{s.t. } & y_p \in \{0, \ldots, m_p-1\}, \quad \forall p \in \{0, \ldots, L-1\} \nonumber \\
             & \sum_{p=0}^{L-1} s_{p,y_p} = \sum_{k=0}^{3L} k \cdot h_k \nonumber \\
             & \sum_{k=0}^{3L} h_k = 1
\end{align}

\subsubsection{Benchmarks and results}

The benchmarks in this work included the 11 protein sequences listed in the work of Fox et al. \cite{fox2021} of SARS-CoV-2, humans, and other organisms (100 to 1300 amino acids), 4 larger protein sequences from humans taken from \cite{uniprot2024} (1000 to 1500 amino acids), and 4 extra-large protein sequences from bacteria and the Caenorhabditis roundworm taken from \cite{uniprot2024} (11000 to 14000 amino acids). Fox et al. \cite{fox2021} benchmarked a subset of 10 of these proteins with a genetic algorithm and D-Wave HQA. For reference, a study in mRNA secondary structure prediction \cite{alevras2024mrna} employed datasets of 10,000 randomly-generated sequences between 6-20 nucleotides (significantly smaller than biologically-relevant sequences) leading to QUBOs of up to 327 variables; a study in peptide-protein docking \cite{brubaker2025quadratic} used a dataset of 6 peptide-protein complexes from the RCSB Protein Data Bank (PDB).
\\
Table \ref{tab:dataset_stats} breaks down the dataset, where average size is the number of QUBO variables for each protein sequence, and density and interconnectivity are defined in Section \ref{sec:qubo_metrics}. The extra-large proteins were only benchmarked with the solvers supporting CP formulations as they performed the best, and a QUBO representation of these proteins was not possible in reasonable memory usage; the table hence does not include the density and interconnectivity for these proteins.
\begin{table}[H]
  \centering
  \footnotesize
  \begin{threeparttable}
    \caption{Dataset statistics}
    \label{tab:dataset_stats}
    \begin{tabularx}{\linewidth}{>{\centering\arraybackslash}X>{\centering\arraybackslash}X>{\centering\arraybackslash}X>{\centering\arraybackslash}X>{\centering\arraybackslash}X>{\centering\arraybackslash}X}
      \toprule
      \textbf{Name}                & \textbf{Dataset size} & \textbf{Avg size} & \textbf{Avg density} & \textbf{Avg IC} \\
      \midrule
      Standard proteins (from \cite{fox2021}) & 11                    & 1540.181         & 0.394               & 0.786                         \\
      Large proteins               & 4                     & 4371.000         & 0.418               & 0.836                         \\
      Extra-large proteins               & 4                     & 45810.500         & N/A              & N/A                         \\
      \bottomrule
    \end{tabularx}
    \normalsize
    \textbf{Avg IC}: average interconnectivity
  \end{threeparttable}
\end{table}
As one can see, the dataset is highly dense and has a high interconnectivity, especially given the GC terms in the objective function. It should be noted that this optimization problem also has a high rank-1 dominance, as the GC terms form a rank-1 QUBO equation and contribute to the majority of the connectivity in the QUBO since the repetition terms (the other quadratic terms) are local to adjacent codons.
\\
The Fujitsu DA v4, various D-Wave HQA solvers, Gurobi (12.0.1), CP-SAT (9.14.6206), HiGHS (1.11.0), and SCIP (9.0) were used to solve the QUBO instances. For D-Wave, the following solvers were applied: Leap Hybrid NL Solver (1.22), Leap Hybrid CQM Solver (1.13), Leap Hybrid BQM Solver (2.2), and the dwave-hybrid workflow solvers Kerberos and Parallel Tempering (PT, custom workflow); the dwave-hybrid solvers utilized the Advantage2 quantum computer version 1.5. The classical MIP/CP solvers were run on a server with 48 cores and 98 GB of RAM as were the classical algorithms of the dwave-hybrid Kerberos and PT HQA solvers. The DA and classical MIP/CP solvers solved all problems to optimality, with the D-Wave Leap CQM HQA achieving near-optimal solutions in all cases. A high-level performance comparison is shown in Table \ref{tab:solver_perf}. Average cost is the average final Hamiltonian cost function value of the solutions found by each solver. Average TTS is the average solve time across all problems, and the standard deviation and maximum of these times are also given. Average QPU usage is the average time the QPU was used by the D-Wave HQA solvers.

\begin{figure}[H]
  \centering
  \includegraphics[width=1.\textwidth]{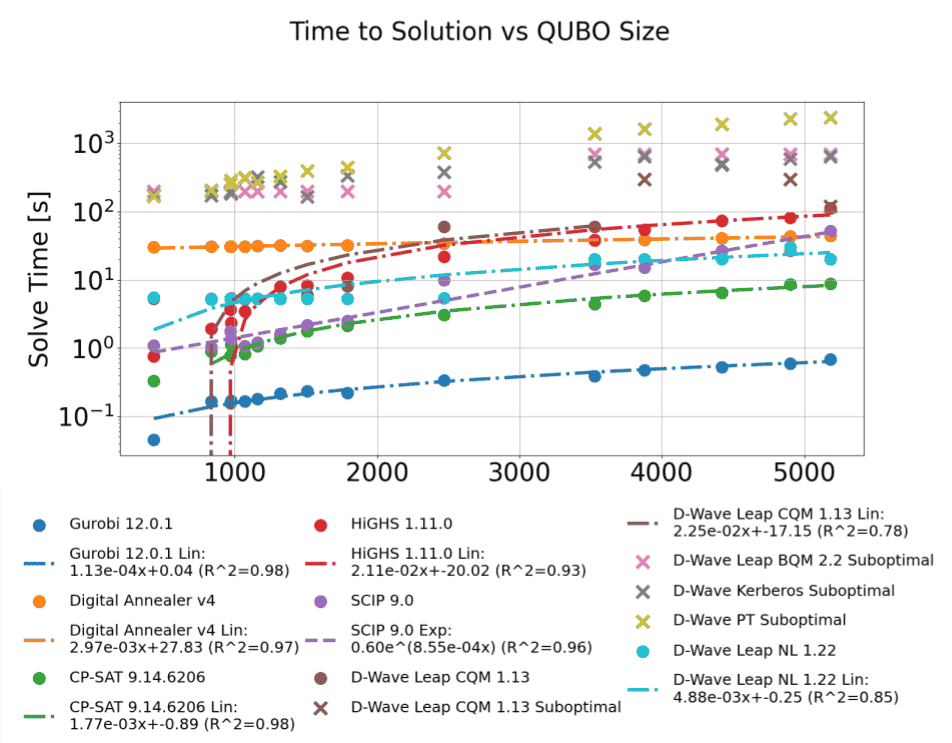}
  \caption{Time to solution plot comparison (Standard and Large Proteins)}
  \label{fig:problem_scaling}
\end{figure}

\begin{table}[H]
  \centering
    \caption{Solver performances comparison (Standard and Large Proteins)}
    \label{tab:solver_perf}
    \begin{tabularx}{\linewidth}{lXXXXX}
      \toprule
      \textbf{Name}    & \textbf{AC} & \textbf{ATTS}     & \textbf{MTTS}     & \textbf{STTS}     & \textbf{AQPU} \\
      \midrule
      Gurobi           & 188.286     & 0.304             & 0.682             & 0.181             & N/A             \\
      CP-SAT           & 188.286     & 3.062             & 8.297             & 2.742             & N/A             \\
      Leap NL HQA      & 188.286     & 10.957            & 30.343            & 8.342             & 0.493           \\
      SCIP             & 188.286     & 10.789            & 52.513            & 14.339            & N/A             \\
      HiGHS            & 188.286     & 28.476            & 113.067           & 34.571            & N/A             \\
      DA               & 188.286     & 34.642            & 44.17             & 4.752             & N/A             \\
      Leap CQM HQA     & 188.312     & 92.888            & 500.653           & 146.862           & 0.077           \\
      Kerberos HQA     & 251.194     & 361.200           & 648.277           & 171.657           & 0.267           \\
      Leap BQM HQA     & 269.282     & 366.639           & 700.002           & 235.677           & 6.410           \\
      PT HQA           & 358.833     & 866.656           & 2384.645          & 787.106           & 0.496           \\
      \bottomrule
    \end{tabularx}
    \caption*{
    \textbf{AC}: average cost; average final solution cost over all problems \\
    \textbf{ATTS [s]}: average time to solution in seconds \\
    \textbf{MTTS [s]}: maximum time to solution in seconds \\
    \textbf{STTS [s]}: standard deviation of time to solution in seconds \\
    \textbf{AQPU [s]}: average QPU usage in seconds}
\end{table}
\noindent
This high-level comparison shows that on the Standard and Large Protein Datasets, the Gurobi, CP-SAT, and Leap NL HQA solvers have the most reliable efficiency in solving to optimality considering average, maximum, and standard deviation time to solution (TTS). This evaluation focuses on overall performance for the Standard and Large Proteins, hence after average cost, average TTS is considered the most important metric followed by maximum and standard deviation TTS. The Leap CQM HQA is unable to obtain optimal solutions for certain problems although finding near-optimal solutions; the unconstrained D-Wave solvers are unable to identify optimal solutions in a reasonable time. An interesting observation is that the D-Wave proprietary Leap Hybrid BQM solver performs worse than the dwave-hybrid workflow Kerberos solver, and the Leap Hybrid BQM solver uses significantly more QPU time than Kerberos.\\
The problem scaling behavior as a function of the number of variables illustrates the performance difference between the solvers in more detail, shown in Figure \ref{fig:problem_scaling}. Data points from Table \ref{tab:solver_perf} that are not shown represent problems that a solver did not obtain the optimal solution for, and have not been included in obtaining the fitted curves. The number of variables in the figure indicates the number of variables for a protein's QUBO representation, meaning the total number of codons across all amino acids.

\begin{figure}[H]
  \centering
  \includegraphics[width=1.\textwidth]{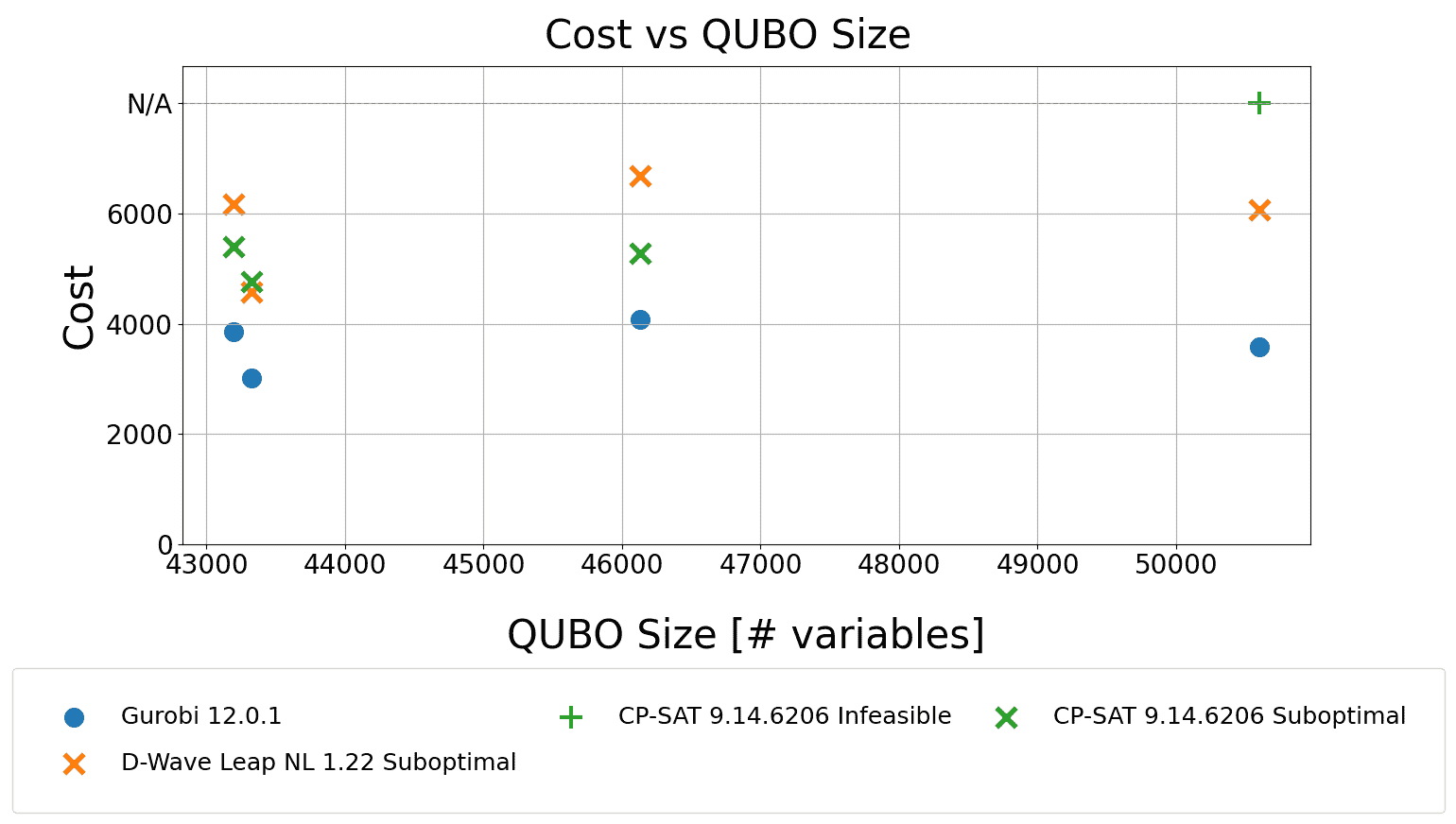}
  \caption{Cost plot comparison (Extra-large Proteins; 60s time limit)}
  \label{fig:xl_cost}
\end{figure}

\noindent
As seen in Figure \ref{fig:problem_scaling} above, the solve time for the Gurobi, CP-SAT, DA, Leap NL HQA, and HiGHS solvers is linear in relation to the problem size, while the solve times for D-Wave and SCIP are exponential. Out of the linear time-scaling solvers, Gurobi has the best scaling behavior (lowest slope), followed by CP-SAT, DA, Leap NL HQA, and then HiGHS. The linear time-scaling advantage of these solvers becomes particularly significant for longer, biologically relevant polypeptide sequences. The Leap CQM HQA was not able to identify optimal solutions with problem sizes closer to 4000 variables and above in reasonable amounts of time, however it achieved near-optimal solutions for all problems. The unconstrained D-Wave solvers are unable to identify any optimal solutions in reasonable amounts of time.
\\
Fox et al. \cite{fox2021} compared D-Wave HQA and quantum approximate optimization algorithm (QAOA) to a genetic algorithm for mRNA codon optimization. The hybrid solver was not able to identify the optimal mRNA solutions for certain protein sequences of only 20 amino acids, and QAOA was not able to do so for even smaller protein sequences. That work applied the D-Wave Leap BQM HQA with the Advantage 1.1 quantum computer, having the one-hot constraints embedded in the cost function; in the current work the D-Wave Leap NL HQA with a CP problem formulation solved all problems in the Standard and Large Protein Datasets to optimality, and the Leap CQM HQA with separated constraints was able to solve problems with under 4000 variables to optimality. Furthermore, the genetic algorithm applied in that work had average solution time of 10.6 minutes for 10 of the full-length proteins (a subset of the dataset applied in this work), a significantly higher amount of time than the solution times for the CP-SAT, Leap NL HQA, DA, HiGHS, and SCIP solvers in the current work that did not exceed 2 minutes for any of the full-length proteins.
\\
The Gurobi, CP-SAT, and D-Wave Leap NL HQA solvers, which had the most efficient solve times for the largest problems among the Standard and Large Protein Datasets, were also benchmarked on the Extra-large Protein Dataset. These solvers were run with a time limit of 60 seconds and the minimum costs obtained in their solutions are compared; Gurobi solved all problems to optimality in an average of 14.064 seconds while CP-SAT and the Leap NL HQA solvers were unable to find optimal solutions within the time limit. The costs obtained by the solvers for the 4 extra-large proteins are shown in Figure \ref{fig:xl_cost}. The number of variables in the figure indicates the total number of codons across all amino acids (equivalent to the number of variables for a protein's QUBO or MILP representation, or the total number of options in the CP formulation).

\noindent
Gurobi outperforms the other solvers on the Extra-large Protein Dataset by solving these problems to optimality under the time limit. The performance of the CP-SAT and Leap NL HQA solvers was comparable, with the CP-SAT outperforming the Leap NL HQA on 2 of the 4 problems, and the Leap NL HQA outperforming CP-SAT on 2 of the 4 problems. For one of the problems in which the Leap NL HQA outperformed CP-SAT, CP-SAT was unable to find a feasible solution in the time limit.

\section{Lessons learned}
This work analyzed the utility of QUBO solvers for chemical reaction network pathway analysis and mRNA codon selection. Tables \ref{tab:problem_mapping}, \ref{tab:connectivity}, and \ref{tab:penalty_structure} summarize the evaluation of these problems under the proposed benchmarking framework, based on the problem structures and quantitative results from Section \ref{sec:use_cases}. The CRN use case represents the CRN pathway analysis problem, and the mRNA use case represents the mRNA codon selection problem.

\begin{table}[H]
  \centering
  \begin{threeparttable}
    \caption{Problem mapping metrics analysis}
    \label{tab:problem_mapping}
    \begin{tabularx}{\linewidth}{|>{\centering\arraybackslash}X|>{\centering\arraybackslash}X|>{\centering\arraybackslash}X|}
      \hline
      \textbf{Use case} & \textbf{Pre-processing} & \textbf{Post-processing} \\
                        & \textbf{matrix to matrix} & \textbf{complexity} \\
      \hline
      CRN & Linear / log scaling & Minimal \\
      \hline
      mRNA & One-to-one scaling & Minimal \\
      \hline
    \end{tabularx}
  \end{threeparttable}
\end{table}

\begin{table}[H]
  \centering
  \begin{threeparttable}
    \caption{Quantitative connectivity analysis}
    \label{tab:connectivity}
    \begin{tabularx}{\linewidth}{|>{\centering\arraybackslash}X|>{\centering\arraybackslash}X|>{\centering\arraybackslash}X|>{\centering\arraybackslash}X|}
      \hline
      \textbf{Use case} & \textbf{Density} & \textbf{Interconnec-tivity} & \textbf{Rank-1 Dominance} \\
      \hline
      CRN & Very low  & Low & N/A \\
      & (linear problem) & & (linear problem)\\
      \hline
      mRNA & Moderate & High & High \\
      \hline
    \end{tabularx}
  \end{threeparttable}
\end{table}

\begin{table}[H]
  \centering
  \begin{threeparttable}
    \caption{Penalty structure analysis}
    \label{tab:penalty_structure}
    \begin{tabularx}{\linewidth}{|>{\centering\arraybackslash}X|>{\centering\arraybackslash}X|>{\centering\arraybackslash}X|}
      \hline
      \textbf{Use case} & \textbf{Constraint Type} & \textbf{Penalty Separation} \\
      \hline
      CRN & Linear & Separated \\
      \hline
      mRNA & One-hot & Separated, combined for D-Wave unconstrained HQA \\
      \hline
    \end{tabularx}
  \end{threeparttable}
\end{table}
\noindent
This work found Gurobi with a linearized MILP formulation was the most efficient for solving the mRNA codon use case, with the CP-SAT and D-Wave NL HQA solvers displaying the next best performance. In addition to these solvers, the Fujitsu DA and HiGHS solved mRNA codon selection to optimality with linear scaling in time to solution. The D-Wave Leap CQM HQA was able to solve problems other than the largest ones to optimality but exhibited exponential scaling in time to solution, and the unconstrained D-Wave HQA solvers were unable to identify optimal solutions. For the CRN pathway analysis use case, classical MIP/CP solvers were able to solve the problem to optimality in reasonable runtimes while the Fujitsu DA was not able to.
\\
The Fujitsu DA is designed to solve combinatorial optimization problems using binary variables. In this work we assessed the DA’s capabilities of also encoding integer variables for solving the reaction network pathway analysis use-case. The integer variables encoding contributed to the performance degradation of the Fujitsu DA.
\\
During the course of this study the following lessons have been learned:
\begin{itemize}
  \item Problems with one-to-one mapping to QUBO can be better suited for quantum and quantum-inspired solvers compared to ones with logarithmic, linear, or superlinear bloating due to binarization.
  \item Linear problems with low density and interconnectivity are often better-suited for classical MIP/CP solvers.
  \item In problems with high density and interconnectivity, when such problems also have high rank-1 dominance classical MIP/CP solvers can potentially solve the problem efficiently with linearizations.
  \item Quantum and quantum-inspired solvers have overhead costs which can lead to more inefficient solve times for smaller problem sizes, but potentially better scaling behavior to larger sizes compared to classical MIP/CP solvers; better scaling behavior of quantum and quantum-inspired solvers was not observed in this work, although the D-Wave Leap NL HQA and DA exhibited linear time scaling for mRNA codon selection.
  \item Constraint-based solvers can offer significant benefits in efficiency in comparison to having to embed penalty terms in a cost function, as seen with the comparison of the constrained and unconstrained D-Wave HQA solvers in this work, and the unconstrained HQA solver usage in the reference literature of the two analyzed use cases \cite{mizuno2024optimal, fox2021}.
\end{itemize}
Furthermore, an important takeaway from optimization problem studies in general is to analyze the differences between mathematical models and real-world problems. For instance, in this work as well as in \cite{fox2021}, a simplified model of mRNA codon optimization was used with three factors (codon usage bias, GC content control, and repeated nucleotide minimization) to evaluate the utility of quantum and quantum-inspired solvers. However, in a realistic and industrial setting, there are many more factors to consider, such as mRNA secondary structure, mRNA folding stability around the ribosome, and more \cite{fox2021, kim1997, brule2017, kudla2009coding}. One may observe that a QUBO-based solver performs better than a classical solver in a simplified model, but may not perform as well in a utility-scale problem formulation. Even in similar applications, such as this mRNA codon optimization problem, mRNA secondary structure as studied by Alevras et al. \cite{alevras2024mrna}, peptide-docking as studied by Brubaker et al. \cite{brubaker2025quadratic}, and protein folding as studied by Romero et al. \cite{romero2025proteinfoldingalltoalltrappedion}, different conclusions have been drawn about the utility of quantum and other QUBO-based solvers due to the nature of the problems and assumptions in the mathematical models.
\\
In mRNA secondary structure prediction, conditional value at risk (CVaR)-based variational quantum eigensolver (VQE) run on IBM quantum computers was able to match structure predictions of the classical CPLEX solver in mRNA sequences up to 42 nucleotides; the unconstrained problem formulation for VQE involved embedding a constraint in the cost function. This work demonstrated an exponential scaling in time to solution with classical solvers however did not assess how quantum solvers scale in solving the problem with over 100 qubits. While the work demonstrates progress in quantum solvers for mRNA structure applications, the problem formulation which involved secondary structure base-pairing did not account for tertiary structure contacts \cite{alevras2024mrna}. Conversely, in the peptide docking problem, a more complex formulation with 3D distance constraints and steric effects, the authors found that mapping the problem to a HOBO binary optimization matrix was expensive in time, and also had to convert from HOBO to QUBO; these factors contributed to their results highlighting that QUBO optimization is not a good fit for this problem as they benchmarked QUBO-based SA against a constraint-programming solver \cite{brubaker2025quadratic}.
Furthermore, for protein folding, a higher-order unconstrained binary optimization (HUBO) formulation was used to model the problem incorporating amino acid interactions (instead of modeling codons and nucleotides for mRNA problems) and terms up to 5th order; the bias-field digitized counterdiabatic quantum optimization (BF-DCQO) algorithm run on IonQ quantum computers found optimal or high-quality near-optimal solutions in all cases \cite{romero2025proteinfoldingalltoalltrappedion}, although Tuziemski et al.\cite{Tuziemski2025} have highlighted shortcomings in how these results have been benchmarked and that statements about quantum advantage are misleading.

\section{Future work}
This work evaluated the utility of quantum and digital annealing solvers for chemical reaction network pathway analysis and mRNA codon selection.
\\
It would be useful to evaluate use cases with different kinds of QUBOs, such as problems with quadratic constraints and dense problems that do not have high rank-1 dominance. Problems with superlinear bloating due to binarization, or HOBO to QUBO mappings would also be interesting benchmarks but potentially more difficult for QUBO solvers.

\section{Conclusions}

This work developed a benchmarking framework for evaluating the utility of quantum annealing, digital annealing, and classical solvers for combinatorial optimization problems formulated as QUBOs with constraints. The framework analyzes problem mapping metrics, quantitative connectivity measures, and penalty structures to systematically assess when quantum and quantum-inspired methods may offer advantages over classical approaches. Through benchmarking two industrially relevant use cases in chemistry and life sciences, namely reaction network pathway analysis and mRNA codon selection, we observed that problem structure influences performance of the different kinds of solvers. For reaction network pathway analysis problems, which exhibit linear/logarithmic bloating in QUBO formulation, low density, and linear constraints, classical solvers (Gurobi, HiGHS, SCIP, CP-SAT) outperformed digital annealing in both accuracy and efficiency. For mRNA codon selection, the problems had one-to-one mapping to QUBO, higher density and interconnectivity, one-hot constraints, however they also had high rank-1 dominance. Gurobi solved all problems (up to 14000 amino acids) the most efficiently; the Gurobi, CP-SAT, Fujitsu DA, D-Wave Leap NL HQA, and HiGHS solvers achieved optimal solutions with linear computational scaling compared to exponential scaling with the D-Wave Leap CQM HQA and SCIP solvers. The D-Wave Leap CQM HQA which natively supports constraints found near-optimal solutions for all problems, and unconstrained D-Wave HQA solvers were unable to identify near-optimal solutions in reasonable amounts of time. In the Extra-large Protein Dataset (11000 to 14000 amino acids), Gurobi exhibited the best performance, and the D-Wave NL HQA solver performed comparably to CP-SAT. These findings highlight the importance of problem-specific evaluation and provide practical guidance for selecting appropriate optimization methods based on QUBO structure and constraint types. 

\section{Acknowledgements}
The authors would like to thank Fujitsu for providing access to the digital annealer during a proof of concept project and the Fujitsu Accelerator Program, as well as Matthieu Parizy for his support during these programs. We would also like to thank QuantumBasel for providing access to D-Wave quantum annealers, and appreciate Julien Baglio for his advice in benchmarking these solvers. Additionally, we would like to thank Helmut Katzgraber (55 North) for his advice in constraint programming and presenting benchmark results.

\bibliography{./arxiv_paper.bib}

\end{document}